%% file: 0_main.tex
\documentclass[acmtog]{acmart}

\usepackage{booktabs} 

\usepackage[ruled]{algorithm2e} 

\SetAlFnt{\small}
\SetAlCapFnt{\small}
\SetAlCapNameFnt{\small}
\SetAlCapHSkip{0pt}

\setcopyright{cc}
\setcctype{by-nc-nd}
\acmJournal{TOG}
\acmYear{2026} \acmVolume{45} \acmNumber{6} \acmArticle{262}
\acmMonth{12} \acmDOI{10.1145/3842552}

\usepackage[capitalize]{cleveref}
\crefname{section}{Sec.}{Secs.}
\Crefname{section}{Section}{Sections}
\Crefname{table}{Table}{Tables}
\crefname{table}{Tab.}{Tabs.}

\usepackage{amsmath}
\usepackage{bbm}

\usepackage{afterpage}
\usepackage{makecell}
\usepackage[table]{xcolor} 
\usepackage{multicol,multirow}

\usepackage{soul}
\usepackage[dvipsnames]{xcolor}

\newcommand{\src}{\mathrm{src}}
\newcommand{\tgt}{\mathrm{tgt}}
\newcommand{\kin}{\mathrm{kin}}
\newcommand{\geo}{\mathrm{geo}}

\begin{document}
\title{Skinned Motion Retargeting via Artifact-driven Kinematic Prior Refinement}

\author{Seokhyeon Hong}
\authornote{Equal contribution.}
\affiliation{%
  \institution{KAIST}
  \country{South Korea}
}
\email{ghd3079@kaist.ac.kr}

\author{Chaelin Kim}
\authornotemark[1]
\affiliation{%
  \institution{KAIST}
  \country{South Korea}
}
\email{chaelin.kim@kaist.ac.kr}

\author{Inseo Jang}
\affiliation{%
  \institution{KAIST}
  \country{South Korea}
}
\email{isjang08@kaist.ac.kr}

\author{Soojin Choi}
\affiliation{%
  \institution{KAIST}
  \country{South Korea}
}
\email{97choisj@kaist.ac.kr}

\author{Junyong Noh}
\affiliation{%
  \institution{KAIST}
  \country{South Korea}
}
\email{junyongnoh@kaist.ac.kr}
\renewcommand\shortauthors{Hong, S. et al}

\begin{teaserfigure}
  \vspace{-0.5em}
  \includegraphics[width=\textwidth]{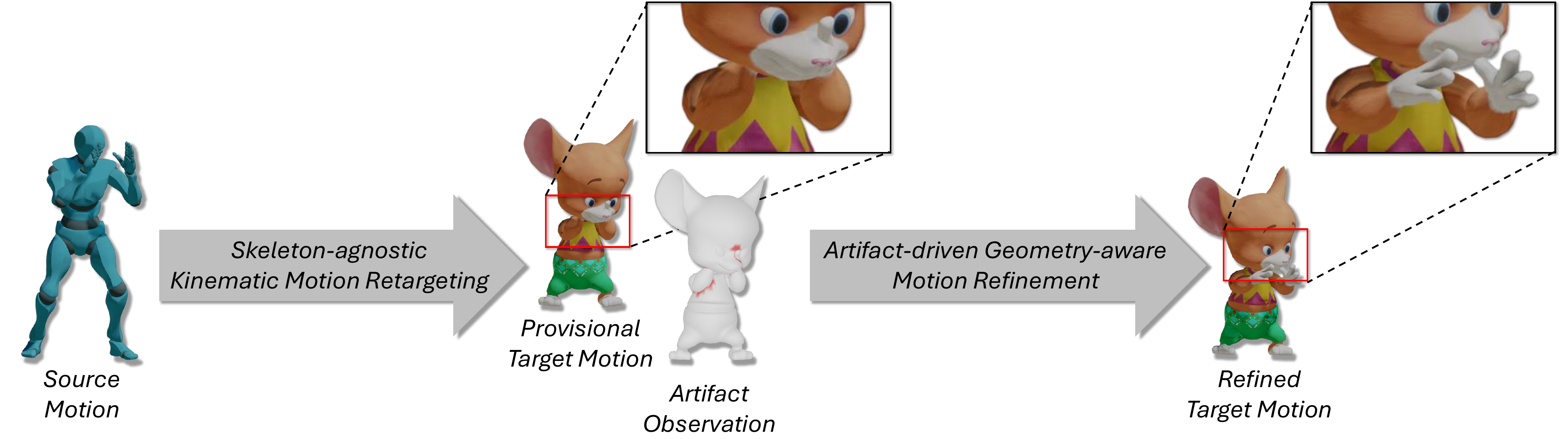}
  \vspace{-2.5em}
  \caption{
  We present a skinned motion retargeting framework via artifact-driven kinematic motion prior refinement.
  Given a provisional motion produced by a transformer-based skeleton-agnostic kinematic motion retargeting module, our method converts posed geometric artifacts into corrective cues in the shared motion embedding.
  }
  \label{fig:teaser}
\end{teaserfigure}

\input{tex/0_abstract}

%
%


%
%


\maketitle

\input{tex/1_intro}
\input{tex/2_related}
\input{tex/3_method}

\input{tex/4_experiments}
\input{tex/5_discussion}
\input{tex/6_conclusion}

\begin{acks}
This work was supported by the National Research Foundation of Korea (NRF) grant funded by the Korea government (MSIT) (RS-2024-00333478).
\end{acks}

\bibliographystyle{ACM-Reference-Format}
\bibliography{bib/0_main}


\newpage
\appendix
\renewcommand{\thetable}{\Alph{table}}
\renewcommand{\thefigure}{\Alph{figure}}
\renewcommand{\theequation}{\Alph{equation}}

\setcounter{table}{0}
\setcounter{figure}{0}
\setcounter{equation}{0}

\input{tex/9_supp}






\end{document}

%% file: tex/0_abstract.tex
\begin{abstract}
Motion retargeting aims to transfer a source motion to target characters with different skeletal structures, proportions, and body shapes.
Although recent neural retargeting methods have improved flexibility across diverse skeletons, target-side geometric artifacts such as self-penetration remain difficult to resolve.
Specifically, existing geometry-aware approaches often rely on fixed skeleton templates or implicit geometry-conditioned prediction, requiring a single network to account for target geometry deformation, detect target-side artifacts, and predict the corresponding correction from target geometry alone, which limits their ability to generalize across diverse skeleton structures and body shapes.
In this paper, we present a geometry-aware motion retargeting framework that explicitly connects artifacts observed in the posed character geometry to motion refinement while preserving the flexibility of skeleton-agnostic neural retargeting.
Our method first learns a motion embedding shared across different skeletons using a transformer-based retargeting autoencoder that transfers motion across arbitrary source--target skeleton pairs.
Building on this kinematic motion prior, we introduce an artifact-driven refinement module that observes self-penetration on the posed target mesh and converts it into a corrective cue through a motion-to-vertex Jacobian.
We further condition motion decoding on target geometry using skinning weight-based joint-aligned geometry features derived from the rest pose mesh.
This design combines explicit target-side artifact reasoning with flexible geometry-aware decoding in a unified framework.
Experiments on both fixed and arbitrary skeleton structure settings show that our method improves kinematic retargeting accuracy and reduces geometric artifacts, producing plausible motions across seen and unseen target characters.
Code is available at \emph{\url{https://seokhyeonhong.github.io/projects/kinematic-refinement/}}.
\end{abstract}

%% file: tex/1_intro.tex
\section{Introduction}
\label{sec:intro}

Character animation is a fundamental element in various computer graphics applications, such as video games and films.
High-quality motion is typically authored through manual keyframing or motion capture, both of which are costly and time-consuming.
To enhance cost efficiency by reusing existing motion data, \textit{motion retargeting} aims to transfer a motion sequence in a source character to diverse target characters with different skeleton structures and body shapes, while preserving the semantic intent of the original motion.

A central challenge in motion retargeting is that a character motion is not defined solely by kinematic properties, such as joint rotations.
In practical character rigs, skeletal motion is ultimately realized through skinning, which deforms the target mesh according to both kinematic articulation and character-specific geometry.
As a result, a retargeted motion that appears plausible in joint space may still produce clear geometric artifacts on the target character, such as self-penetration.
Therefore, a robust retargeting framework requires both kinematic fidelity and geometric compatibility with the target character.

Recent neural motion retargeting methods have made substantial progress through learning-based formulations that employ neural networks as retargeting solvers.
In particular, neural kinematic retargeting has shown that motion transfer can be performed efficiently in a feed-forward manner by learning motion patterns that generalize across diverse characters, thereby bypassing iterative and computationally expensive optimization process~\cite{villegas2018nkn, lim2019pmnet, aberman2020san, lee2023same, hu2023pose, zhang2024unified}.
Beyond computational efficiency, these methods have reduced reliance on explicit source--target joint correspondences by learning motion representations shared across different skeletons~\cite{aberman2020san, lee2023same}, leading to flexibility across diverse skeleton pairs with different numbers of joints or skeletal hierarchies, even being generalizable to target skeletons unseen during training.
These advances highlight the main advantages of neural retargeting in terms of efficiency, flexibility, and generalizability across diverse characters.

However, these advantages have not yet been fully realized in geometry-aware retargeting.
Although recent methods improve visual plausibility by incorporating target geometry~\cite{zhang2023r2et, yang2025star, ye2024meshret, zhang2024smt}, many still remain limited in flexibility and generalizability that neural retargeting has achieved at the kinematic level.
In particular, geometry-aware correction is often built on fixed skeleton templates or shared geometry structure assumptions, such as anchors, which makes it difficult to handle characters with different joint counts or asymmetric skeletons.
Moreover, many learning-based approaches rely on the rest pose geometry alone and expect a single model to implicitly infer the entire correction process at once: (i) detecting where artifacts will occur, (ii) identifying which parts of the motion are responsible, and (iii) predicting how the motion should be modified to resolve them.
As a result, the key advantages of neural retargeting are not yet fully realized in the geometry-aware setting, which helps explain why optimization-based correction methods continue to remain relevant despite their iterative test-time cost and sensitivity to objective design~\cite{martinelli2024moma, cheynel2025reconform, guo2025ultrafast}.

In this paper, we propose a learning-based geometry-aware motion retargeting framework designed to preserve the main advantages of neural retargeting in a geometry-aware setting.
Our main idea is to decompose geometry-aware motion retargeting into three atomic responsibilities--kinematic motion retargeting, artifact observation, and motion refinement--instead of expecting a single model to solve the entire problem implicitly.
This decomposition allows each module to focus on a well-defined role while preserving flexibility across diverse skeleton structures and meshes with arbitrary numbers of vertices.

To enable flexible retargeting across diverse characters within a unified framework, we first train a transformer-based autoencoder~\cite{vaswani2017attention} that learns a motion embedding shared across different skeletons from arbitrary source--target motion pairs without relying on pre-defined skeleton structures.
A key motivation for this design of the kinematic module is that motion retargeting depends not only on local joint relations but also on global coordination across the full-body, which self-attention between joints can model more directly.
This shared motion embedding, together with its decoder, serves as a reliable motion prior that captures motion semantics transferable across diverse characters.

Given the provisional retargeted motion, we then introduce an artifact-driven geometry-aware motion refinement module to mitigate target-side geometric artifacts.
Inspired by optimization-based methods that use geometric gradients to guide motion updates, we convert artifacts observed in the posed character geometry into a first-order corrective cue using a motion-to-vertex Jacobian, which describes how changes in the shared motion embedding affect posed geometry.
Importantly, we do not use this cue to directly optimize any variables through iterative refinement; instead, we use it as a feed-forward estimation of motion prior correction.
Furthermore, we update the motion embedding instead of per-joint motion features, such as joint rotations, thereby preserving flexibility across diverse skeleton structures.
Finally, because the corrective cue is derived from artifacts observed on the posed target geometry, it provides an explicit target-specific signal about where the artifact occurs and how the motion embedding should change to alleviate it.
This reduces reliance on inferring corrections solely from the rest pose geometry, and thus generalizes better to unseen characters and geometric configurations such as body shapes.

Although the rest pose geometry alone is not sufficient for effective geometry-aware motion retargeting, it still provides useful target-specific geometric context for the decoding process.
For example, characters with thicker arms should keep their arms farther from the torso than thinner characters under similar motions.
To achieve this without relying on fixed templates or shared geometry parameterizations, we introduce a skinning weight-based skeleton enrichment scheme that injects character-specific geometric information into the target skeleton representation in a joint-aligned manner.
By using skinning weights to associate geometry with skeletal joints, our enrichment provides the decoder with flexible geometric context while preserving dimensional compatibility across diverse characters.
With these components combined, our method reduces target-side geometric artifacts while remaining robust to unseen characters with diverse skeletons and body shapes.

In summary, our contributions are as follows:
\begin{itemize}
    \item We present a learning-based geometry-aware motion retargeting framework that combines a shared skeleton-agnostic motion prior with explicit motion-to-vertex Jacobian-based artifact-driven refinement.
    \item We introduce a transformer-based kinematic retargeting module for skeleton-agnostic motion retargeting, which preserves flexibility across diverse skeleton structures while capturing global coordination and fine-grained joint details.
    \item We propose a skinning weight-based skeleton enrichment scheme that converts character-specific geometry into joint-aligned geometric context for motion decoding while preserving flexibility across diverse skeletons.
\end{itemize}

%% file: tex/2_related.tex
\section{Related Work}
\label{sec:related}

\subsection{Kinematic Motion Retargeting}
Motion retargeting is the process of transferring motion between different characters while ensuring that the target character follows the source motion accurately and naturally.
Traditional methods primarily relied on space-time optimization~\cite{gleicher1998retarget}, inverse kinematics~\cite{choi2000online, shin2001computer}, and constraint-based solver~\cite{tak2005physically}.
To accommodate topological differences across characters, \citeN{monzani2000using} introduced anatomic binding into a shared intermediate skeleton.
Recent approaches employ neural networks trained on large-scale datasets to achieve efficient inference and generalization across diverse skeletons~\cite{villegas2018nkn, lim2019pmnet, aberman2020san, lee2023same, hu2023pose, zhang2024unified}.
NKN~\cite{villegas2018nkn} proposed an unsupervised motion retargeting framework that utilizes a cycle consistency loss without relying on source-target paired motion datasets between characters.
PMnet~\cite{lim2019pmnet} further improved the quality of retargeted motion by learning pose and overall motion separately.
To accommodate homeomorphic skeletons within a shared latent space, SAN~\cite{aberman2020san} introduced character-specific skeleton-aware operations that compress skeletons into a shared primal skeleton, followed by a decoder specific to each target character.
SAME~\cite{lee2023same} further enabled skeleton-agnostic motion retargeting by constructing a motion embedding shared by arbitrary skeletons so that a single decoder can effectively retarget motion across a variety of target characters.
Our method extends these kinematic retargeting approaches by augmenting a skeleton-agnostic retargeting backbone with explicit target-side geometric reasoning for artifact-aware motion refinement.

\subsection{Geometry-aware Motion Retargeting}
Beyond transferring kinematic motion, geometry-aware motion retargeting methods aim to improve the realized target motion at the mesh level, such as reducing self-penetration or preserving contact relationships.
This line of work is motivated by the observation that kinematically plausible motions may still produce unrealistic surface-level artifacts after skinning.
Accordingly, prior methods have incorporated geometric conditions into retargeting either to explicitly repair mesh artifacts through optimization or to learn geometry-compatible motion generation.

For penetration avoidance, optimization-based methods typically begin with a provisional retargeting result, detect self-penetration on the posed target mesh, and then iteratively update motion parameters to reduce the observed artifacts~\cite{lyard2008motion, martinelli2024moma}.
Recent learning-based methods instead address this process in a feed-forward manner by conditioning motion retargeting on target geometry.
R2ET~\cite{zhang2023r2et} introduced distance fields to train a shape-aware module that softly discourages penetration artifacts, while STaR~\cite{yang2025star} extended this work through spatio-temporal consistency into penetration-aware motion retargeting.
SMT~\cite{zhang2024smt} leveraged a pre-trained vision-language model to incorporate high-level semantic information alongside penetration avoidance.

Another line of studies focuses on contact-aware retargeting, where the objective is to preserve meaningful geometric interactions.
In these studies, interactions are often modeled through pre-defined geometric representations, such as anchors or pair-wise geometric relations, whose relative configurations are preserved.
This category includes self-body interactions~\cite{liu2018surface, basset2019contact, cheynel2023sparse, cheynel2025reconform, villegas2021contact, choi2026skinned}, interactions with the environment~\cite{ho2013motion, choi2023online, jin2025interfacerays}, or geometric relationships between multiple characters~\cite{jang2024geometry, jin2018aura}.

Our method is most closely related to learning-based penetration-aware retargeting, while making a different design choice in how geometry participates in a feed-forward motion refinement.
In our framework, target geometry serves not only as conditioning information, but also as a source of explicit artifact observations from posed character geometry after provisional retargeting, which are translated into a corrective cue through a motion-to-vertex Jacobian, along with character-specific geometry injection through skinning-based alignment.
In this way, our method combines explicit target-side artifact reasoning with geometry-aware decoding, while preserving the flexibility to handle arbitrary skeletal structures and body shapes that previous geometry-aware approaches often struggled with.

%% file: tex/3_method.tex
\section{Method}
\label{sec:method}
Given a source motion $M_\src$ associated with a source skeleton $S_\src$, our goal is to retarget the motion to an arbitrary target skeleton $S_\tgt$ along with its associated geometry $G_\tgt$.
The target skeleton may differ from the source in skeletal structures, including the number of joints, skeletal hierarchy, and bone lengths, while the target geometry may contain an arbitrary number of vertices.
Our framework is designed to handle this structural diversity in both kinematic and geometric variables within a unified retargeting pipeline.
In the following sections, we first describe the data representations used in our framework, including skeleton, motion, and geometry~(\cref{sec:method_data}).
We then present a transformer-based autoencoder that learns a shared motion embedding for skeleton-agnostic kinematic retargeting~(\cref{sec:method_kin}).
Building on this pre-trained motion prior, we introduce a geometry-aware correction module that converts artifacts observed in the posed target geometry into explicit corrective cues in the shared motion embedding and conditions decoding on target-specific geometry through skinning weight-based skeleton enrichment~(\cref{sec:method_geo}).


\subsection{Data Representation}
\label{sec:method_data}

\paragraph{Skeleton Data}
We represent a character skeleton $S$ as follows:
\begin{equation}
    S=\{ g, o \},
\end{equation}
where $g\in\mathbb{R}^{N_J\times3}$ and $o\in\mathbb{R}^{N_J\times3}$ represent the global joint positions and parent-relative joint offsets in the rest pose (i.e., T-pose), respectively.
Here, $N_J$ denotes the number of joints.
To align coordinate conventions across characters, we follow the pre-processing procedure of SAME~\cite{lee2023same}, which applies a reset transformation to all joint rotations such that they are set to identity in the rest pose.

\paragraph{Motion Data}
A motion sequence is denoted by $M^{1:N_T}$, where $N_T$ is the number of frames.
The pose representation at frame $t$ is defined as follows:
\begin{equation}
    M^t=\{q^t, p^{t-1}, p^t, \Delta p^{t}, r^t, c^t \},
\end{equation}
where $q^t\in\mathbb{R}^{N_J\times6}$ denotes the parent-relative joint rotation represented using the continuous 6D rotation representation~\cite{zhou2019continuity}.
The terms $p^t\in\mathbb{R}^{N_J\times3}$ and $\Delta p^t\in\mathbb{R}^{N_J\times3}$ denote the joint position and velocity relative to the character's facing frame~\cite{holden2017phase}, respectively.
The root feature $r^t\in\mathbb{R}^{4}$ is defined by 2-dimensional translational velocity on the ground plane~($xz$-plane) and 1-dimensional rotational velocity along the up vector~($y$-axis), along with 1-dimensional height of the root joint.
Finally, $c^t\in\mathbb{R}^{N_J\times1}$ is a binary ground contact label computed using height and velocity thresholds~\cite{lee2002interactive}.

\paragraph{Geometry Data}
We represent the character geometry $G$ in the rest pose and its skinning information as follows:
\begin{equation}
    G=\{ v, n, W \},
\end{equation}
where $v\in\mathbb{R}^{N_V\times3}$, $n\in\mathbb{R}^{N_V\times3}$, and $W\in\mathbb{R}^{N_V\times N_J}$ denote the vertex positions, vertex normals, and skinning weight matrix, respectively.
Here, $N_V$ denotes the number of vertices.


\input{fig/same_limitation}
\subsection{Kinematic Motion Retargeting}
\label{sec:method_kin}
Our kinematic retargeting module transfers a source motion sequence $M_\src$ defined on a source skeleton $S_\src$ to a target skeleton $S_\tgt$ even when the two skeletons differ in skeletal structures, such as the number of joints, skeletal hierarchy, and bone lengths.
This formulation is closely related to SAME~\cite{lee2023same}, which constructs a motion representation shared across diverse characters.
However, we observe that its retargeted motions may inaccurately reproduce fine-grained joint details, particularly the orientations of end effector joints that convey important motion semantics such as pointing and gaze direction as shown in~\cref{fig:same_limitation}.
This limitation may partly stem from its graph convolutional architecture: although SAME aggregates joint features over the skeleton using graph attention~\cite{velivckovic2017graph}, its message passing remains restricted to local graph neighborhoods.
Consequently, information between distant joints must be propagated through multiple layers, which may have difficulty in modeling long-range dependencies~\cite{he2023generalization}.
In contrast, pose-level semantics often depend on coordinated relationships among distant joints that are not local in the skeletal hierarchy.

To this end, we adopt a transformer-based autoencoder with global self-attention over all joint tokens, which allows each joint token to directly interact with every other joint token within a single layer while learning a shared motion representation from arbitrary source--target skeleton pairs without relying on a fixed skeletal structure.
Specifically, we formulate kinematic retargeting as an autoencoder architecture that maps source motion on an arbitrary skeleton to a shared motion embedding, and reconstructs it on a target skeleton that may contain a different skeletal structure.
Each joint is represented as a token that combines both static skeletal and dynamic motion information, while a learnable motion token aggregates motion semantics.
The decoder then conditions this shared representation on the target skeleton to produce retargeted motion.
This design provides a skeleton-agnostic motion prior while improving reconstruction accuracy of fine-grained joint details by modeling global relationships across skeletal joints through self-attention.

\input{fig/kin_overview}
\subsubsection{Architecture}
As shown in~\cref{fig:kin_overview}, for each frame $t$, we construct a set of source joint tokens by combining static skeletal features and dynamic motion features as follows:
\begin{equation}
    x_\src^t = [S_\src, M_\src^t]_\mathrm{ch} \in\mathbb{R}^{N_J \times 26},
\label{eqn:x_src}
\end{equation}
where $[\cdot,\cdot]_\mathrm{ch}$ denotes concatenation of the skeletal and motion features for each joint along the channel dimension.
Because the root feature $r_\src^t$ in $M_\src^t$ is defined at the character level, we assign it only to the root joint and pad zeros for all non-root joints to match the token dimensionality, while this padding is omitted in the notation for brevity.
Furthermore, we omit positional encoding based on token order, because skeletal semantics should not depend on joint ordering.
Instead, each joint token is identified by its corresponding features in $S$, which provide joint-specific rest pose information that distinguishes the joint from the others.

Each source token is then processed by a stack of transformer encoder layers $\mathcal{T}_\src(\cdot)$ together with a learnable motion token $z$:
\begin{gather}
    z^t_\kin = \mathcal{T}_\src \left( [z, x_\src^t]_\mathrm{tok} \right)_z,
\end{gather}
where $[\cdot,\cdot]_\mathrm{tok}$ denotes concatenation along the token dimension, while $(\cdot)_z$ denotes selecting the output embedding corresponding to the learnable motion token $z$.
Here, $z^t_\kin\in\mathbb{R}^{D}$, denoting $D$ as the latent dimension, serves as the shared motion embedding for frame $t$.
Note that, instead of obtaining the motion embedding by directly pooling joint features, we learn it through a dedicated motion token that interacts with all source joint tokens through self-attention.
This design separates joint-level feature encoding from aggregating global motion semantics, allowing the embedding to adaptively collect the information relevant for retargeting, while remaining compatible with different skeletal structures.
In this sense, the motion token plays a role similar to class tokens in transformer architectures for high-level summarization~\cite{caron2021dino, devlin2019bert}, enabling input-dependent aggregation that adaptively gathers motion-relevant information from individual joint tokens, unlike fixed pooling based on a predefined aggregation rule.

Given the static features of the target skeleton $S_\tgt$, we represent them as a set of target joint tokens, which are processed by the target-side transformer $\mathcal{T}_\tgt(\cdot)$ along with the shared motion embedding:
\begin{equation}
\begin{aligned}
    \hat M_\tgt^t &= [\hat q^t_\tgt, \hat r^t_\tgt, \hat c^t_\tgt] \\
                  &= \mathcal{T}_\tgt \left([z^t_\kin, S_\tgt]_{\mathrm{tok}}\right)_\mathrm{joint},
\end{aligned}
\end{equation}
where $(\cdot)_\mathrm{joint}$ denotes selecting the output embeddings corresponding to the joint tokens.
Similar to~\cref{eqn:x_src}, the root feature $\hat r_\tgt^t$ is associated only with the root joint, and the non-root predictions are ignored, while we omit this distinction in the notation for brevity.
Through self-attention, each target joint token can directly attend both to the shared motion embedding and to all other target joints.
This design allows the motion semantics encoded in $z_\kin^t$ to be distributed adaptively over the target skeleton, while remaining compatible with target skeletons with different skeletal structures.
Further details are provided in the supplementary material~(Sec.~A.1).

\input{fig/artifact_overview}

\subsubsection{Training}
\label{sec:method_kin_training}
Primarily inspired by SAME~\cite{lee2023same}, the loss function used to train the kinematic retargeting transformer is defined as follows:
\begin{equation}
    \mathcal{L}_\kin=\sum_{t=1}^T \big(\mathcal{L}_\mathrm{rec}^t + \mathcal{L}_\mathrm{smooth}^t + \mathcal{L}_\mathrm{con}^t + \mathcal{L}_\mathrm{emb}^t\big).
\end{equation}

\paragraph{Reconstruction Loss}
This term encourages accurate reconstruction of   motion features, including joint rotations, positions, and root features, as follows:
\begin{equation}
    \begin{aligned}
        \mathcal{L}_\mathrm{rec}^t
        =
        &\lambda_q
        \lVert q_\tgt^t-\hat q_\tgt^t \rVert_2^2
        +
        \lambda_p
        \lVert p_\tgt^t- \hat p_\tgt^t \rVert_2^2 \\
        &+
        \lambda_r
        \lVert r_\tgt^t-\hat r_\tgt^t \rVert_2^2.
    \end{aligned}
\end{equation}
The predicted joint position is computed as follows:
\begin{equation}
    \hat p_\tgt^t=\mathrm{FK}(S_\tgt, \hat q_\tgt^t, \hat h_\tgt^t),
\end{equation}
where $\mathrm{FK}(S,q,h)$ computes forward kinematics using the skeleton $S$, joint rotations $q$, and root height $h$.

\paragraph{Smoothness Loss}
This term encourages temporal coherence by matching the linear joint velocities between adjacent frames and penalizing excessive acceleration changes:
\begin{equation}
    \begin{aligned}
        \mathcal{L}_\mathrm{smooth}^t
        =
        &\lambda_\mathrm{vel}
        \lVert
        \Delta p_\tgt^t
        -
        \Delta \hat p_\tgt^t
        \rVert_2^2
        +
        \lambda_\mathrm{jerk}
        \lVert
        (\hat a_\tgt^t - \hat a_\tgt^{t-1}) / \Delta t
        \rVert_2^2,
    \end{aligned}
\end{equation}
where the velocity and acceleration are defined as follows:
\begin{gather}
\Delta \hat p_\tgt^t
= \frac{(\hat p_\tgt^t - \hat p_\tgt^{t-1})}{\Delta t},
\quad
\hat a_\tgt^t
= \frac{(\Delta \hat p_\tgt^t  - \Delta \hat p_\tgt^{t-1} )}{\Delta t}.
\end{gather}

\paragraph{Contact Loss.}
This term encourages accurate ground contact prediction and reduces foot sliding and ground penetration, which is defined as follows:
\begin{equation}
    \begin{aligned}
    &\mathcal{L}_\mathrm{con}^t
    =
    \lambda_c
    \lVert c_\tgt^t-\hat c_\tgt^t \rVert_2^2
    +
    \lambda_{cv}
    \lVert c_\tgt^t \cdot \Delta \hat p_\tgt^t \rVert_2^2 \\
    &+
    \lambda_\mathrm{slide}
    \lVert \operatorname{clamp} (1-\frac{\hat y_\tgt^t}{y_\mathrm{thres}}, 0, 1) \cdot \Delta \hat p_\tgt^t \rVert_2^2
    +
    \lambda_\mathrm{gp}
    \lVert \min(\hat y_\tgt^t,0) \rVert_2^2,
    \end{aligned}
\end{equation}
where $y$ denotes the $y$-component from joint positions while $y_\mathrm{thres}$ denotes the height threshold used to detect ground contact .
The first term supervises contact labels, the second suppresses joint velocity during contact, the third reduces sliding for joints near the ground, and the last term penalizes ground penetration.

\paragraph{Contrastive Embedding Loss}
This term encourages the motion embedding to discriminatively preserve motion semantics while reducing skeleton-specific information.
As described in~\cref{sec:training_details}, each training sample consists of a short source--target motion window, and mini-batches are constructed by grouping multiple such pairs together.
Within each mini-batch, we treat embeddings from samples that share the same underlying source motion as positive pairs, while treating the others as negative pairs.
The contrastive loss is defined as:
\begin{equation}
\begin{aligned}
\mathcal{L}_z^t=
&\frac{1}{|\mathcal{P}|}
\sum_{(i,j)\in\mathcal{P}}
\lVert z_{\kin, i}^t - z_{\kin, j}^t \rVert_2^2
+ \\
&\frac{1}{|\mathcal{N}|}
\sum_{(i,j)\in\mathcal{N}}
\max(0,m-\lVert z_{\kin, i}^t - z_{\kin, j}^t \rVert_2)^2,
\end{aligned}
\end{equation}
where $\mathcal{P}$ and $\mathcal{N}$ denote the positive and negative pair sets, respectively, and $m$ is the margin.
Unlike the embedding consistency loss in previous retargeting studies~\cite{lee2023same, aberman2020san}, which only pulls corresponding source-target embeddings together, our contrastive formulation also pushes embeddings of different motions apart, encouraging a more discriminative motion embedding space.

\subsection{Artifact-driven Geometry-aware Motion Refinement}
\label{sec:method_geo}

Given the pre-trained kinematic retargeting module, we obtain a provisional target motion that preserves the source motion semantics at the skeleton level.
However, because this prediction is made primarily in joint space, the posed target geometry obtained by deforming the skinned mesh with the predicted pose may exhibit geometry-level artifacts, such as self-penetration.
Correcting such artifacts is challenging because they are observed on the target geometry, while the retargeting model must ultimately modify kinematic features.

Accordingly, we introduce an artifact-driven geometry-aware motion refinement module that explicitly observes artifacts in the posed target geometry and translates them into corrective cues for motion refinement, rather than relying on implicit geometry-conditioned prediction solely from the rest pose geometry.
In addition, the correction should remain compatible with diverse target characters whose meshes may contain an arbitrary number of vertices and whose skeletons may have different structures.
To this end, our module decomposes refinement into a shared motion embedding correction and target-specific geometry conditioning.
The artifact-driven Jacobian cue is used to predict an update in the shared motion embedding rather than directly modifying target joint angles, thereby keeping the correction within the skeleton-agnostic motion prior and allowing it to be realized through the pre-trained decoder.
By describing how the posed target geometry changes with respect to the motion embedding, the Jacobian cue provides a motion-aware direction for artifact correction.
However, it does not fully determine the appropriate correction for a specific character, because the required adjustment also depends on the target body shape, such as torso thickness and limb size.
We therefore derive geometric features from the static target geometry to enrich the target skeleton tokens with character-specific shape information.
Accordingly, the motion embedding update modifies the motion in response to the observed artifact, while the geometry enrichment allows the decoder to realize the corrected motion according to the target character.

\subsubsection{Artifact-driven Corrective Cue}
As shown in the left side of~\cref{fig:artifact_overview}, we first deform the target character's geometry using linear blend skinning~(LBS) given the provisional retargeted motion:
\begin{equation}
(\hat v_\tgt^t, \hat n_\tgt^t) = \operatorname{LBS}\left(v_\tgt, n_\tgt, W_\tgt, \operatorname{FK}(S_\tgt, \hat q_\tgt^t, \hat h_\tgt^t)\right),
\label{eqn:lbs}
\end{equation}
where $\hat v^t_\tgt$ and $\hat n^t_\tgt$ denote the posed target vertices and normals at frame $t$, respectively.
Subsequently, we follow STaR~\cite{yang2025star} to detect self-penetration by treating the vertices associated with each arm and each leg as query vertices, while using those of the remaining body regions as reference vertices.
For each query vertex, we identify its nearest reference vertex and compute their relative displacement:
\begin{equation}
    d^t = \hat v_{\tgt,\mathrm{ref}}^{t} - \hat v_{\tgt,\mathrm{query}}^t.
\label{eqn:disp}
\end{equation}
Using the corresponding reference normal $\hat n_{\tgt,\mathrm{ref}}^t$, we measure the signed penetration signal as follows:
\begin{equation}
    s^t = \max (\hat n_{\tgt,\mathrm{ref}}^t \odot d^t, 0),
\label{eqn:pen_signal}
\end{equation}
where $\odot$ represents the dot product.
Intuitively, $d^t$ specifies how a penetrating query vertex should move locally to reduce the observed artifact, while $s^t$ measures the extent to which displacement is aligned with the penetration direction.
Implementation details for efficient parallel computation across heterogeneous vertex counts are provided in the supplementary material~(Sec.~A.2).

To relate these vertex-level artifact signals, including $d^t$ and $s^t$, to motion refinement, we translate the local corrective displacement into the shared motion embedding space.
Specifically, the shared motion embedding $z_\kin^t$ is converted to posed target geometry through decoding, forward kinematics, and skinning, all of which are fully differentiable.
This allows us to define the motion-to-vertex Jacobian:
\begin{equation}
    J^t_{v, z} = \frac{\partial \hat v^t_\tgt}{\partial z_\kin^t},
\end{equation}
which describes how infinitesimal changes in the motion embedding affect the posed target vertices.
Using this Jacobian, we map the local corrective displacement $d^t$ defined on penetrating vertices into a first-order corrective cue:
\begin{equation}
    u^t = (J^t_{v,z})^\top (\mathbb{I}(s^t>0) \cdot d^t),
\end{equation}
where $\mathbb{I}(\cdot)$ denotes the indicator function.
This cue indicates how the shared motion embedding should change in a first order to reduce the observed penetration.

\subsubsection{Geometry-conditioned Refinement}
We do not directly apply $u^t$ for iterative optimization updates.
Instead, we use it as an explicit input to a feed-forward refinement module that predicts an embedding update $\Delta z^t_\kin$, which approximates a motion correction that can resolve penetration in the final output:
\begin{gather}
    \Delta z^t_\kin = \phi_\mathrm{corr}([z_\kin^t, \operatorname{Norm}_2 (u^t)]_\mathrm{ch}),\\
    z^t_\geo = z_\kin^t + \Delta z^t_\kin,
\end{gather}
where $\operatorname{Norm}_p(x)=x/(\lVert x \rVert_p + \epsilon)$ denotes $L_p$ normalization using a small constant $\epsilon$ for numerical stability.
Here, $\phi_\mathrm{corr}(\cdot)$ is an MLP that predicts an embedding residual from the current shared motion embedding and the artifact-driven corrective cue, which is normalized for training stability.
To effectively preserve the learned motion prior, we initialize the last layer of $\phi_\mathrm{corr}(\cdot)$ with zeros so that the correction $\Delta z^t_\kin$ starts with zero at the beginning of training, which improves training stability in the early stage~\cite{zhang2023adding}.

While the artifact-driven corrective cue refines the shared motion embedding, effective retargeting also requires the decoding process itself to be conditioned on target-specific geometry.
Specifically, the same corrected motion embedding may need to be realized differently depending on character-specific shape, such as limb thickness or body proportions.
Therefore, the decoder $\mathcal{T}_\tgt(\cdot)$ should access target-specific geometric context in addition to the corrected motion embedding.
In addition, such geometry conditioning should remain compatible with arbitrary target characters whose meshes may contain different numbers of vertices and whose skeletons may have different structures.

To satisfy this requirement, we use skinning weight as a natural association between geometric vertices and skeletal joints.
Unlike fixed templates or shared geometry parameterizations, skinning weights directly encode how each mesh vertex is influenced by each joint, and therefore provide a flexible mechanism for converting vertex-level geometry into a joint-aligned representation.
Specifically, as shown in the right side of~\cref{fig:artifact_overview}, we first extract vertex-wise geometry features from the rest pose geometry of the target character using a shared MLP $\phi_\mathrm{geo}(\cdot)$, and then aggregate them into a joint-aligned geometric context $j_\tgt$ through the normalized transpose of skinning weight matrix:
\begin{equation}
    j_\tgt = \operatorname{Norm}_{1}(W_\tgt^\top)\, \phi_{\geo}([v_\tgt, n_\tgt]_{\mathrm{ch}}).
\end{equation}
We then project this geometric context into a residual bias to enrich the target skeleton tokens using an MLP $\phi_\mathrm{enr}(\cdot)$:
\begin{equation}
    b_\tgt = \phi_{\mathrm{enr}}(j_\tgt).
\end{equation}
Notably, we use rest pose vertices rather than posed target vertices to enrich the target skeleton tokens with geometric information.
Specifically, because the target skeleton tokens encode character-specific information independent of the current pose, we use rest pose geometry to provide pose-independent shape context, such as body proportions and limb thickness, for the target skeleton representation.
In contrast, posed geometry depends on the provisional motion and is already used to derive the artifact-driven corrective cue, which guides the prediction of the motion-embedding update $\Delta z_{\mathrm{kin}}^t$.
This separation maintains distinct roles for static target-specific geometry and pose-dependent artifact information during motion refinement.

Similar to $\phi_\mathrm{corr}(\cdot)$, the last layer of $\phi_\mathrm{enr}(\cdot)$ is initialized with zeros to improve training stability in the early stage.
Finally, the decoder produces the refined target motion by taking the corrected motion embedding and the geometry-enriched target skeleton tokens as input:
\begin{equation}
\begin{aligned}
    \tilde M_\tgt^t &= [\tilde q^t_\tgt, \tilde r^t_\tgt, \tilde c^t_\tgt] \\
                    &= \mathcal{T}_\tgt \left([z^t_\geo, S_\tgt + b_\tgt]_{\mathrm{tok}}\right)_\mathrm{joint}.
\end{aligned}
\end{equation}
Note that, during geometry-aware training, all kinematics-related modules introduced in~\cref{sec:method_kin} are kept frozen, and only the newly introduced modules are optimized, namely $\phi_\mathrm{corr}(\cdot)$, $\phi_\mathrm{geo}(\cdot)$, and $\phi_\mathrm{enr}(\cdot)$.

\subsubsection{Training}
We train the geometry-aware modules using the following objective:
\begin{equation}
    \mathcal{L}_\mathrm{geo}=\sum_{t=1}^T \big(\mathcal{L}_\mathrm{rec}^t + \mathcal{L}_\mathrm{smooth}^t + \mathcal{L}_\mathrm{con}^t + \mathcal{L}_\mathrm{pen}^t + \mathcal{L}_\mathrm{sparse}^t\big),
\label{eqn:loss_geo}
\end{equation}
where $\mathcal{L}_{\mathrm{rec}}^{t}$, $\mathcal{L}_{\mathrm{smooth}}^{t}$, and $\mathcal{L}_{\mathrm{con}}^{t}$ are the reconstruction, smoothness, and contact losses introduced in~\cref{sec:method_kin_training}, respectively, while we apply geometry-aware predictions $\tilde{(\cdot)}$ instead of kinematic predictions $\hat{(\cdot)}$.
Because the kinematic backbone is kept frozen during geometry-aware training, these terms preserve the original retargeting behavior and discourage unnecessary deviation from the provisional motion.
On top of them, we introduce a penetration loss and a sparsity regularization loss for geometry-aware refinement.

\paragraph{Penetration Loss}
We first obtain the updated vertex positions and normals using~\cref{eqn:lbs} but with $\tilde q^t_\tgt$ and $\tilde h^t_\tgt$, and derive a new signed penetration signal $\tilde s^t$ following~\cref{eqn:disp,eqn:pen_signal}.
The penetration loss is then defined as follows:
\begin{equation}
    \mathcal{L}_\mathrm{pen}^t = \lambda_\mathrm{pen} \frac{\sum_i \tilde s_i^t}{\sum_i \mathbb{I}(\tilde s_i^t > 0) + \epsilon}.
\end{equation}
This term penalizes self-penetration on the posed target mesh and encourages the corrected motion to reduce the observed artifacts.

\paragraph{Sparsity Regularization Loss}
The sparsity regularization loss is defined as
\begin{equation}
    \mathcal{L}_\mathrm{sparse}^t = \lambda_\mathrm{sparse} \left(\lVert \Delta z_\kin^t \rVert_2^2 + \lVert b_\tgt \rVert_2^2\right),
\end{equation}
which regularizes both the motion embedding update and the geometry-enrichment residual to remain small.
This encourages conservative refinement around the provisional target motion while preserving the learned kinematic motion prior.

%% file: fig/same_limitation.tex
\begin{figure}[t]
    \centering
    \includegraphics[width=\linewidth]{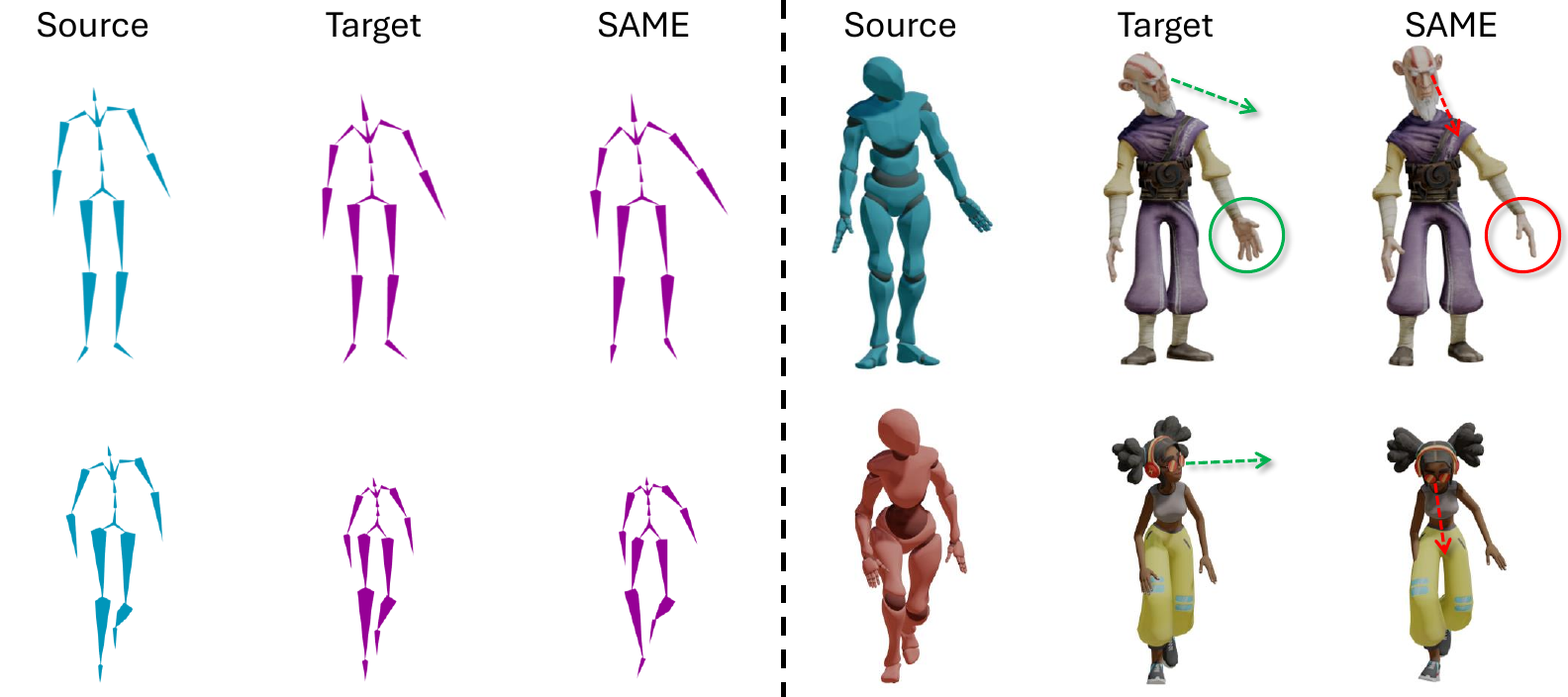}
    \vspace{-2em}
    \caption{Limitation of SAME in preserving accurate joint rotations.}
   \label{fig:same_limitation}
   \vspace{-1em}
\end{figure}

%% file: fig/kin_overview.tex
\begin{figure}[t]
    \centering
    \includegraphics[width=\linewidth]{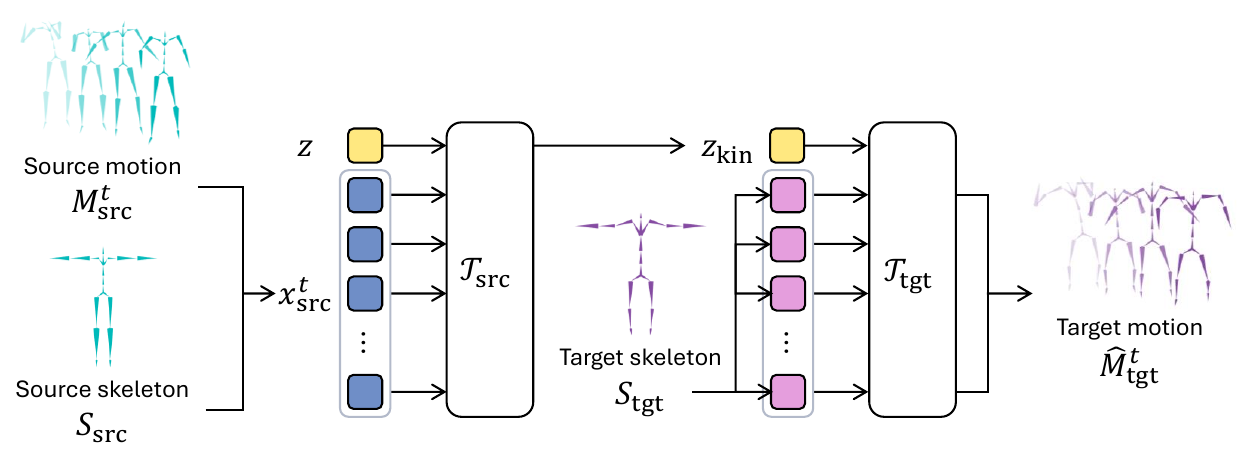}
    \vspace{-2em}
    \caption{Overview of the kinematic retargeting transformer.}
   \label{fig:kin_overview}
    \vspace{-1em}
\end{figure}

%% file: fig/artifact_overview.tex
\begin{figure*}[t]
    \centering
    \includegraphics[width=\linewidth]{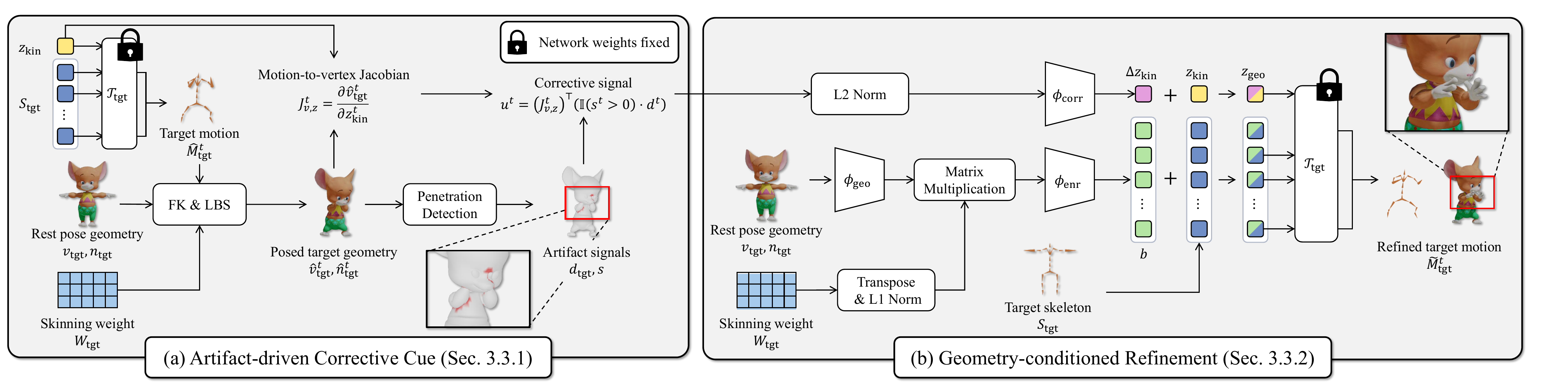}
    \vspace{-2em}
    \caption{Overview of the artifact-driven correction mechanism.}
   \label{fig:artifact_overview}
    \vspace{-1em}
\end{figure*}

%% file: tex/4_experiments.tex
\input{tab/kin_retargeting}

\section{Experiments}
\label{sec:experiments}

\subsection{Implementation Details}
\subsubsection{Training Details}
\label{sec:training_details}
The training dataset for the kinematic autoencoder was constructed from public motion datasets, including ACCAD~\cite{accad}, LaFAN1~\cite{harvey2020robust}, Mixamo~\cite{mixamo}, PFNN~\cite{holden2017phase}, SFU~\cite{sfu}, and TotalCapture~\cite{trumble2017total}.
Using MotionBuilder, we retargeted each motion sequence to five target skeletons randomly selected from a pool of 91 skeletons, yielding 13.85 hours of source--target motion pairs in total, and source--target pairs except those from Mixamo were adopted from the training dataset of SAME.
For the artifact-driven geometry-aware motion refinement module, we used the Mixamo subset of the kinematic training dataset, along with 7 skinned Mixamo characters.
Both the kinematic autoencoder and geometry-aware refinement modules are trained with the Adam optimizer~\cite{kingma2014adam}.
The kinematic autoencoder is trained for 300 epochs with a batch size of 512~(64 samples with 8 consecutive frames).
We used an initial learning rate of $5\times10^{-4}$, which is decayed by a factor of $0.99$ after every epoch.
The geometry-aware refinement module is trained for 200 epochs with a batch size of 192~(24 samples with 8 consecutive frames), using an initial learning rate of $1\times10^{-5}$ with the same epoch-wise decay schedule.
Training the kinematic autoencoder and geometry-aware refinement module required 7.5 and 11 hours, respectively, on a single NVIDIA RTX 6000 Ada GPU.
Please refer to Sec.~A.3 in the supplementary material for additional implementation details, such as loss weights.

\subsubsection{Comparison Methods}
We compare our method against four representative baselines: MotionBuilder~\cite{motionbuilder}, SAME~\cite{lee2023same}, R2ET~\cite{zhang2023r2et}, and STaR~\cite{yang2025star}.
MotionBuilder serves as a production-standard retargeting baseline, and it is also used to generate the ground truth target motions.
SAME is a strong kinematic baseline that introduces skeleton-agnostic motion retargeting through a shared motion embedding independent of the underlying skeleton structure.
R2ET is a representative geometry-aware motion retargeting method that incorporates target geometry to mitigate geometric artifacts, while STaR further models spatio-temporal consistency for smoother retargeting results, serving as a strong recent baseline for geometry-aware motion retargeting.
Additionally, because R2ET and STaR are defined only for the fixed skeleton setting where source and target share the same number of joints, we evaluate them only in the fixed structure setting in both kinematic and geometry-aware motion retargeting.
To ensure a fair comparison, we followed the pre-processing pipeline of each method, including its coordinate system conventions.

\subsubsection{Metrics}
We evaluate motion retargeting from two complementary perspectives.
The kinematic metrics assess how accurately and naturally the retargeted motion matches the ground truth target motion, while the geometry metrics assess how plausibly the realized target mesh avoids self-penetration after skinning.
For kinematic evaluation, we report four metric scores: joint rotation~(JR) that measures rotational error in radians, root translation~(RT) that measures positional error of the root trajectory on the ground plane, joint position~(JP) that measures positional error with respect to the global coordinate in centimeters, and foot sliding~(FS) that measures velocities during contact.
All kinematic metrics are computed against the ground truth target motion generated by MotionBuilder~\cite{motionbuilder}.
For geometry-aware evaluation, we additionally report penetration ratio (PR), defined as the percentage of query vertices detected as penetrating, and penetration depth (PD), defined as the normal-direction penetration magnitude computed by the dot product between the query-to-reference displacement and the corresponding reference normal.


\input{tab/geometry_template_retargeting}
\subsection{Quantitative Results}
\label{sec:exp_quan}
\subsubsection{Kinematic Motion Retargeting}
\label{sec:exp_quan_kin}
We first evaluate the kinematic motion retargeting performance to validate the effectiveness of our proposed transformer-based backbone.
In the evaluation, we exclude MotionBuilder from the comparison because it is used to the generate ground truth motions, and would therefore trivially yield zero error.
In addition, because STaR is designed as an integrated geometry-aware retargeting framework without an explicitly separable kinematic-only counterpart, we compare in this section against SAME and a kinematic-only version of R2ET, denoted as R2ET-Kin.

To assess both reconstruction accuracy and generalization across all compared methods, we evaluate on both seen and unseen target characters, while all motion sequences used for evaluation were excluded from training.
For each seen--unseen split, we consider two experiment settings: a fixed structure setting, where source and target characters share the same skeleton structure but differ only in bone lengths, and an arbitrary structure setting, where target characters may have different numbers of joints, skeletal hierarchies, and bone lengths.
The fixed structure setting includes 4 seen and 10 unseen target characters, while the arbitrary structure setting includes 91 seen and 91 unseen target characters.
For each target character, we evaluate on 95 motion sequences from Mixamo that are often used for kinematic retargeting evaluation in prior studies~\cite{lee2023same, aberman2020san, villegas2018nkn}, which consist primarily of locomotion-oriented motions but with a rich set of target skeletons.
This motion set is particularly suitable for kinematic evaluation because it provides consistent test cases for measuring dynamic root trajectories, local joint motions, and diverse foot contact patterns.

As shown in~\cref{tab:kin_retargeting}, our method achieved the best overall kinematic retargeting performance across both fixed and arbitrary settings, for both seen and unseen target characters.
Compared with SAME, our method consistently improved all metrics in all evaluation settings.
R2ET-Kin achieved the best JR scores in both seen and unseen cases of the fixed setting, which we attribute to its copy-based formulation.
Specifically, R2ET is designed for the fixed skeleton setting, where source joint rotations are first copied to the target and the network predicts only a residual on top of this initialization.
Under this formulation, joint rotations are already strongly biased by direct copying, and thus even a small residual correction can yield a favorable JR score.
In contrast, our method predicts retargeted motion directly in a skeleton-agnostic motion embedding while not relying on a copy-based initialization.
Despite not being the best on JR in the fixed template setting, our method achieved substantially lower scores in RT, JP, and FS across both seen and unseen characters while remaining close to the best results in JR.
Overall, these results show that the proposed transformer-based backbone provides a strong and generalizable kinematic foundation for motion retargeting, with consistent gains across both seen and unseen target characters while remaining applicable to arbitrary target skeletons beyond fixed template settings.

\subsubsection{Geometry-aware Motion Retargeting}
\label{sec:exp_geo}
We evaluate geometry-aware motion retargeting in the fixed structure setting using both seen and unseen target characters while following the same unseen motion protocol as in the kinematic evaluation.
For the same reason as in the kinematic evaluation, we do not report MotionBuilder in the kinematic metrics and instead include only its penetration metrics.
Additionally, R2ET and STaR are limited by a fixed skeleton structure, we limit this comparison to the fixed structure setting for compatibility.
For evaluation, we used 7 seen and 4 unseen skinned characters from Mixamo, along with 102 motion sequences provided by MotionBuilder database~\cite{motionbuilder}.
Compared with Mixamo motion sequences, this set contains not only locomotions but also a broader range of actions involving self-body interactions, making it a more challenging benchmark for evaluating geometry-aware retargeting.

The experimental results are shown in~\cref{tab:geometry_template_retargeting}.
Overall, baselines without explicit geometry conditioning, including MotionBuilder and SAME, exhibited larger penetration artifacts despite remaining competitive on some kinematic metrics, such as FS.
R2ET reduced penetration compared to SAME and achieved the best in JR, which is consistent with its copy-based residual formulation as described in~\cref{sec:exp_quan_kin}.
However, this advantage in JR did not translate into better overall motion quality, as indicated by its relatively large RT, JP, and FS scores.
STaR further improved the penetration-related metrics over R2ET in the seen character setting by evaluating penetrations against a broader set of body regions.
Specifically, while R2ET mainly focuses on penetrations between limbs and a limited set of body regions~(i.e., head and torso), STaR accounts for penetrations between each limb and other remaining body parts, along with spatio-temporal geometry modeling.
However, this improvement came with degraded kinematic performance and limited generalization as indicated by its worse penetration performance than R2ET in the unseen character setting.
In contrast, our method achieved the best performance on RT, JP, PR, and PD by a clear margin, while achieving competitive kinematic performance.
Overall, these results show that explicitly translating posed target-side artifacts into motion-consistent corrective updates is more effective than relying on the rest pose geometry conditioning alone.

\input{tab/userstudy}
\subsubsection{User Study}
To evaluate the naturalness of the motions retargeted by different methods in human perception, we conducted a user study.
We recruited 20 participants~(11 males and 9 females; ages 22--36), and they were presented with a total of 50 videos~(10 source--target motion pairs $\times$ 5 methods).
They were asked to rate their preferences in the following aspects: (i) semantics preservation~(SP), (ii) artifact avoidance~(AA), and (iii) overall quality~(OQ).
Each question was randomly shuffled to avoid ordering bias and was rated on a 5-point Likert scale, with 5 being the best.

As shown in~\cref{tab:user_study}, our method achieved the highest scores in all three criteria, along with the best overall average.
MotionBuilder showed competitive performance in SP, but its lower scores in AA and OQ suggest that kinematic plausibility alone was not sufficient to produce the most convincing retargeted results.
SAME also preserved motion semantics effectively, but its lack of explicit mesh awareness limited its perceptual quality, especially in AA and OQ.
R2ET achieved favorable results among the baselines, indicating that geometry-aware refinement improves perceptual realism, but it still remained consistently below our method.
STaR received the lowest scoress in all three metrics, which is consistent with its weaker kinematic performance in the quantitative evaluation.
These results suggest that our method provides the most favorable perceptual balance between kinematic fidelity and geometric plausibility, leading to the most favorable results in human perception.


\input{fig/kin_qual}
\input{fig/geo_qual}
\input{tab/ablation_geometry}
\input{fig/ablation_geo}

\subsection{Qualitative Results}
\subsubsection{Kinematic Motion Retargeting}
\label{sec:exp_qual_kin}
To qualitatively compare the kinematic retargeting results of our method and the baselines, we visualize a source character alongside different target characters at the same frame in~\cref{fig:kin_qual}.
While both SAME and R2ET-Kin produced plausible motions at a coarse kinematic level, fine-grained motion details were often weakened.
In particular, SAME showed limited reconstruction of end effector orientations, especially at the head and hands.
The gains over SAME, which employs a graph convolutional backbone, suggest that modeling global dependencies between joints using a transformer backbone helps preserve fine-grained motion details in kinematic retargeting.
R2ET-Kin preserved the overall pose structure more faithfully than SAME, but its applicability is limited to fixed template skeleton structures.
In contrast, our transformer-based kinematic retargeting module preserved fine local joint details while remaining flexibility across diverse skeletons.
These qualitative results are consistent with the quantitative results in~\cref{sec:exp_quan_kin}, indicating that the proposed kinematic transformer provides a stronger motion prior for retargeting across diverse characters.
For animated results, please refer to the supplementary video.

\subsubsection{Geometry-aware Motion Retargeting}
We visualize geometry-aware retargeting results in~\cref{fig:geo_qual}.
Both MotionBuilder and SAME frequently exhibited visible self-penetration, showing that kinematically plausible retargeting alone is insufficient to ensure geometric plausibility after skinning.
R2ET produced cleaner results in many cases and reduced minor penetrations, but it still failed to prevent large penetrations in more challenging poses, such as those involving both hands held close together.
Furthermore, R2ET often exhibited severe jittering artifacts in cases involving large penetrations on characters with exaggerated body shapes~(see supplementary video).
STaR attempted stronger geometry-aware correction, but often failed to preserve the original motion faithfully, leading to noticeable degradation in overall motion quality.
In contrast, our method resolved self-penetration more reliably while preserving the intended motion semantics, even in challenging cases involving strong self-body interactions under large shape variations.
These results align with the quantitative evaluation and show that our artifact-driven motion refinement framework achieves a better balance between motion fidelity and geometric plausibility with lower penetration artifacts.
Animated results are provided in the supplementary video.


\input{tab/ablation_kin_embedding}
\input{tab/ablation_contrastive}

\subsection{Ablation Study}
\subsubsection{Artifact-driven Geometry-aware Refinement Components}
To evaluate the contribution of each geometry-aware refinement component, we compare five variants: (a) static rest pose geometry conditioning alone using $\phi_\mathrm{geo}(\cdot)$ and $\phi_\mathrm{enr}(\cdot)$ without skinning weights; (b) the Jacobian-based motion embedding update $\Delta z_\kin^t$ alone; (c) their direct combination; (d) static geometry with skinning-based alignment using $\operatorname{Norm}_1(W_\tgt^\top)$ in addition to the setting (a) to construct joint-aligned geometry features; and (e) the full model.
Because these variants ablate components of geometry-aware retargeting framework, we evaluate them under the same dataset and metrics for geometry-aware retargeting described in~\cref{sec:exp_geo}.
For variants (a) and (c), we extract a single global geometry embedding by applying max pooling over the vertex-wise features, which is then shared across all joints, rather than constructing joint-wise geometry features.
As shown in~\cref{tab:ablation_geometry}, the full model achieved the lowest PR and PD in both seen and unseen character settings while also maintaining strong kinematic performance.

The comparison also clarifies the role of each component.
Static geometry alone provides character-specific shape information, but is less effective in reducing penetration when used without explicit artifact-driven motion refinement, as reflected by its relatively high PD and PR.
The Jacobian-only variant improved kinematic metric scores, especially in the unseen character setting because the Jacobian cue captures how posed geometry changes with respect to the motion embedding, providing a motion-aware direction for artifact correction that rest pose geometry cannot provide.
However, it did not substantially reduce geometric artifacts, as reflected by its weaker PR and PD scores.
This suggests that motion sensitivity alone is insufficient, because the model must also know the target body shape to determine where the motion should be redirected and by how much considering its body shape.
Meanwhile, combining static geometry with the Jacobian-based motion embedding update improved the penetration-related metrics over either component alone, but the kinematic scores became less favorable, especially in the unseen character setting.
This indicates that global geometry conditioning is not the most effective way to combine character shape information with kinematic motion, particularly for unseen characters.
Introducing the skinning weight-based alignment improved the balance between kinematic fidelity and geometric plausibility by converting rest pose geometry into joint-aligned features.
When all components were combined, the model attained the best penetration reduction while keeping kinematic errors within a competitive range.
Overall, these results support our full design, where the rest pose geometry, joint-aligned geometry features through skinning weights, and Jacobian-guided motion refinement work together to reduce penetration while preserving kinematic accuracy.

These findings are also consistent in qualitative results shown in~\cref{fig:ablation_geo}.
Using only static geometry or only the Jacobian-based corrective cue can alleviate shallow penetrations to some extent, but neither was sufficient to resolve them completely, and both were largely ineffective for deeper penetrations.
Combining static geometry with the Jacobian cue, or using static geometry with skinning-based alignment, yielded more effective correction and reduced penetration artifacts, but noticeable penetrations still remain.
In contrast, when all components were used together, the model produced the most reliable correction, resolving both shallow and deep penetrations while better preserving the semantics of the source motion.
Consequently, these qualitative ablation results support that static geometry, joint-aligned geometry conditioning through skinning weights, and artifact-driven corrective cues play complementary roles in geometry-aware retargeting.

\subsubsection{Kinematic Motion Embedding}
We conducted two experiments to demonstrate the effectiveness of a learnable motion embedding token and the contrastive embedding objective.
To validate the effectiveness of using a learnable motion embedding token, which encourages adaptive aggregation, we compared it against fixed pooling over the joint features while omitting the motion token.
As shown in~\cref{tab:ablation_kin_embedding}, attention-based information change between the motion token and skeletal features resulted in consistently better reconstruction accuracy in RT and JP while demonstrating tolerable differences in JR and FS.
These results support the design choice of using learnable motion embedding token against traditional pooling-based aggregation.

To evaluate the effectiveness of the embedding objective, we compare three variants: (a) no embedding loss~(i.e., $\mathcal{L}_z$), (b) a pull-only embedding loss that aligns embeddings of the same motion, following previous studies~\cite{lee2023same, aberman2020san}, and (c) our contrastive embedding loss.
As shown in~\cref{tab:ablation_contrastive}, the contrastive variant achieved the best JR in all four evaluation settings and the best JP in three out of four settings while showing only marginal differences in RT compared to the best results.
These gains indicate that a more discriminative embedding improves the quality of motion transfer not only within a fixed skeletal structure but also across structurally different target skeletons, along with generalization to unseen target characters.
The pull-only variant achieved the lowest RT in all four settings and remained competitive on JP, showing that aligning embeddings of the same motion is already beneficial for retargeting.
However, the scores were consistently worse than the contrastive variant on JR, and also underperformed on JP in three out of four settings, indicating its lower accuracy at the kinematic level.
Dropping the embedding loss yielded the best FS in all four settings, but the gap is very small, while the performance of no embedding loss variant was clearly worse on JR and JP.
Overall, these results support our design choice of using a contrastive embedding objective, showing that preserving discrimination between different motions beyond cross-character consistency yields a motion representation with strong generalization for motion retargeting.
We also provide additional experiments demonstrating applications of the shared motion embeddings in Sec.~B of the supplementary material.

\subsubsection{Comparison to Optimization}
\input{tab/optim}
\input{fig/optim}
To evaluate the effectiveness of learning the geometry-conditioned refinement module, we compared our feed-forward framework with gradient-based optimization, which directly optimizes $\Delta z_{\text{kin}}$ and $b$.
Specifically, we initialize both $\Delta z_{\text{kin}}$ and $b$ to zero and jointly optimize them by minimizing $\mathcal{L}_{geo}$ specified in~\cref{eqn:loss_geo}.
We used the Adam optimizer with a learning rate of $1 \times 10^{-3}$ for 100 iterations, while keeping the remaining loss and optimization settings identical to those used for training our geometry-aware refinement module.
We additionally report the average inference time~(IT) per sample in seconds, computed by dividing the total elapsed time by the number of evaluated samples. 

\cref{tab:optim} presents the quantitative comparisons.
Although the optimization achieved the best scores on individual metrics as it iteratively refines penetration artifacts, our learned refinement offers a practical trade-off between refinement quality and computational efficiency.
Specifically, our method is $95.25\times$ faster in Fixed-SC, and $107.85\times$ faster in Fixed-UC.
In addition, \cref{fig:optim} provides complementary observations that are not fully captured by the frame-wise metrics.
Both methods removed most of the visible penetration artifacts, and our method preserved the semantic characteristics of the source motion more faithfully while achieving comparable geometry correction.
Overall, the learned geometry-conditioned refinement module produces stable and semantically consistent refinements without expensive per-instance optimization, providing a substantial advantage in inference time.

\subsubsection{Mesh Resolution Sensitivity}
The training characters have approximately 2K--7K vertices, while the unseen test characters used in our experiments contain up to 20K vertices.
To examine sensitivity to mesh resolution, we decimated character meshes to have approximately 5K vertices and compared the results with those obtained using the original meshes.
As shown in \cref{tab:ablation_decim}, the original and decimated meshes yielded comparable performance in both the Fixed-SC and Fixed-UC settings, with only minor variations across the metrics.
These results suggest that our method generalizes across mesh resolutions, although vertex sampling density can still influence the measured geometry-related scores.
\input{tab/ablation_decim}

\subsection{Additional Results}
\input{fig/ood_mixamo}
\input{fig/ood}
To further demonstrate the generalizability of our framework beyond the unseen characters used in~\cref{sec:exp_geo}, we visualize additional retargeting results on several target characters from Mixamo.
As shown in~\cref{fig:ood_mixamo}, our method produced penetration-free results while preserving the semantics of the source motion.
Notably, even when the source motion itself contained visible penetrations, such as torso--arm, hand--hand, and hand--torso intersections, these artifacts were effectively alleviated in the retargeted results.

We also evaluate our method on more challenging out-of-distribution characters.
As shown in~\cref{fig:ood}, these results include one Mixamo character with an asymmetric arm skeleton and three characters from different datasets, namely LaFAN1~\cite{harvey2020robust}, SMPL~\cite{loper2015smpl}, and RigNet-v1~\cite{xu2020rignet}.
Furthermore, as shown in~\cref{fig:ood2}, we present a more challenging example in which the asymmetric skeleton is used as the source character while the target skeletons remain out-of-distribution characters.
Even in this case, our method was able to reconstruct a plausible target motion while mitigating penetration artifacts.
Overall, these results suggest that our proposed framework remains flexible and effective beyond the character distributions and skeleton structures used in the main evaluation.

\input{fig/self_pen_recovery}
As an extension of our method, we apply our method to penetration recovery through self-reconstruction.
In this setting, the source and target characters are identical, and thus the task reduces to reconstructing the original motion while correcting self-penetration artifacts.
As shown in~\cref{fig:self_pen_recovery}, our method successfully removed visible penetrations while preserving the original motion semantics and overall pose dynamics.
This result highlights that the proposed artifact-driven geometry-aware motion refinement is not limited to cross-character retargeting, but can also serve as a practical motion refinement mechanism for repairing penetrated motions on a single character.

%% file: tab/kin_retargeting.tex
\begin{table*}[t]
\centering
\caption
{
    Quantitative results on kinematic motion retargeting.
    * indicates that we employed only the kinematics retargeting component of the model, which was originally designed for geometry-aware retargeting.
    \textbf{Bold} indicates the best result, while SC and UC represent seen character and unseen character, respectively.
    R2ET-Kin is measured only for the fixed skeleton setting.
}
\vspace{-1em}
\resizebox{\linewidth}{!}
{
    \begin{tabular}{c|cccc|cccc|cccc|cccc}
        \toprule
        \multirow{2}{*}{Methods}
        & \multicolumn{4}{c|}{Fixed-SC}
        & \multicolumn{4}{c|}{Fixed-UC}
        & \multicolumn{4}{c|}{Arbitrary-SC}
        & \multicolumn{4}{c}{Arbitrary-UC} \\
        & JR~$\downarrow$ & RT~$\downarrow$ & JP~$\downarrow$ & FS~$\downarrow$ 
        & JR~$\downarrow$ & RT~$\downarrow$ & JP~$\downarrow$ & FS~$\downarrow$ 
        & JR~$\downarrow$ & RT~$\downarrow$ & JP~$\downarrow$ & FS~$\downarrow$ 
        & JR~$\downarrow$ & RT~$\downarrow$ & JP~$\downarrow$ & FS~$\downarrow$  \\

        \midrule
        
        SAME
        & 0.2845 & 5.9872 & 11.0881 & 0.0196
        & 0.2754 & 6.0979 & 10.6295 & 0.0104
        & 0.2805 & 5.4221 & 10.1841 & 0.0130
        & 0.2789 & 4.9642 & 9.1355 & 0.0143 \\

        R2ET-Kin*
        & \textbf{0.0917} & 6.9634 & 10.3788 & 0.1618
        & \textbf{0.1096} & 3.7629 & 7.5617 & 0.1414
        & \multicolumn{4}{c}{N/A}
        & \multicolumn{4}{c}{N/A} \\

        \rowcolor{black!16}
        Ours-Kin*
        & 0.1561 & \textbf{2.9568} & \textbf{5.2375} & \textbf{0.0063}
        & 0.1565 & \textbf{2.9869} & \textbf{5.0958} & \textbf{0.0026}
        & \textbf{0.1570} & \textbf{2.9294} & \textbf{4.8443} & \textbf{0.0036}
        & \textbf{0.1579} & \textbf{2.5759} & \textbf{4.5406} & \textbf{0.0089} \\
        
        \bottomrule
    \end{tabular}
}
\vspace{-1em}
\label{tab:kin_retargeting}
\end{table*}

%% file: tab/geometry_template_retargeting.tex

            
            
            
            

\begin{table*}[t]
    \centering
    \caption
    {
        Quantitative results on geometric motion retargeting with a fixed template skeleton shared by both source and target characters.
        \textbf{Bold} indicates the best result, while SC and UC represent seen character and unseen character, respectively.
    }
    \vspace{-1em}
        \begin{tabular}{c|cccccc|cccccc}
            \toprule
            \multirow{2}{*}{Methods}
            & \multicolumn{6}{c|}{Fixed-SC}
            & \multicolumn{6}{c}{Fixed-UC} \\
            & JR~$\downarrow$ & RT~$\downarrow$ & JP~$\downarrow$ & FS~$\downarrow$ & PR~$\downarrow$ & PD~$\downarrow$
            & JR~$\downarrow$ & RT~$\downarrow$ & JP~$\downarrow$ & FS~$\downarrow$ & PR~$\downarrow$ & PD~$\downarrow$ \\

            \midrule
            MotionBuilder
            & - & - & - & - & 1.4981 & 1.6037
            & - & - & - & - & 0.5142 & 1.4066 \\
            
            SAME
            & 0.2688 & 10.0614 & 13.9958 & 0.0388 & 1.4270 & 1.6409
            & 0.2492 & 8.8761 & 12.1477 & \textbf{0.0156} & 0.3791 & 1.3280 \\
            
            R2ET
            & \textbf{0.1210} & 11.2236 & 15.5949 & 0.1533 & 1.1313 & 1.5134
            & \textbf{0.1084} & 10.4534 & 13.7412 & 0.0888 & 0.2364 & 1.1317 \\
            
            STaR
            & 0.2146 & 17.5529 & 28.7375 & 0.2756 & 0.8298 & 1.5058
            & 0.2010 & 15.8851 & 26.6064 & 0.2739 & 0.2453 & 1.2868 \\
            
            \rowcolor{black!16}
            Ours
            & 0.1633 & \textbf{4.8692} & \textbf{7.6974} & \textbf{0.0365} & \textbf{0.6434} & \textbf{1.4311}
            & 0.1556 & \textbf{7.4929} & \textbf{10.4406} & 0.0393 & \textbf{0.1674} & \textbf{1.0191} \\
            
            \bottomrule
        \end{tabular}
    \vspace{-1em}
    \label{tab:geometry_template_retargeting}
\end{table*}

%% file: tab/userstudy.tex
\begin{table}[t]
    \centering
    \caption{User study results. \textbf{Bold} indicates the best.}
    \vspace{-1em}
        \begin{tabular}{c|ccc|c}
            \toprule
            Methods & SP & AA & OQ & Average \\
            \midrule
            MotionBuilder
            & 3.965 & 3.115 & 3.420 & 3.500 \\

            SAME
            & 3.685 & 3.325 & 3.400 & 3.470 \\

            R2ET
            & 3.875 & 3.420 & 3.545 & 3.613 \\

            STaR
            & 2.805 & 2.620 & 2.465 & 2.630 \\

            \rowcolor{black!16}
            Ours
            & \textbf{4.275} & \textbf{4.385} & \textbf{4.295} & \textbf{4.318} \\
            \bottomrule
        \end{tabular}
    \label{tab:user_study}
    \vspace{-1em}
\end{table}

%% file: fig/kin_qual.tex
\begin{figure*}[t]
    \centering
    \includegraphics[width=0.9\linewidth]{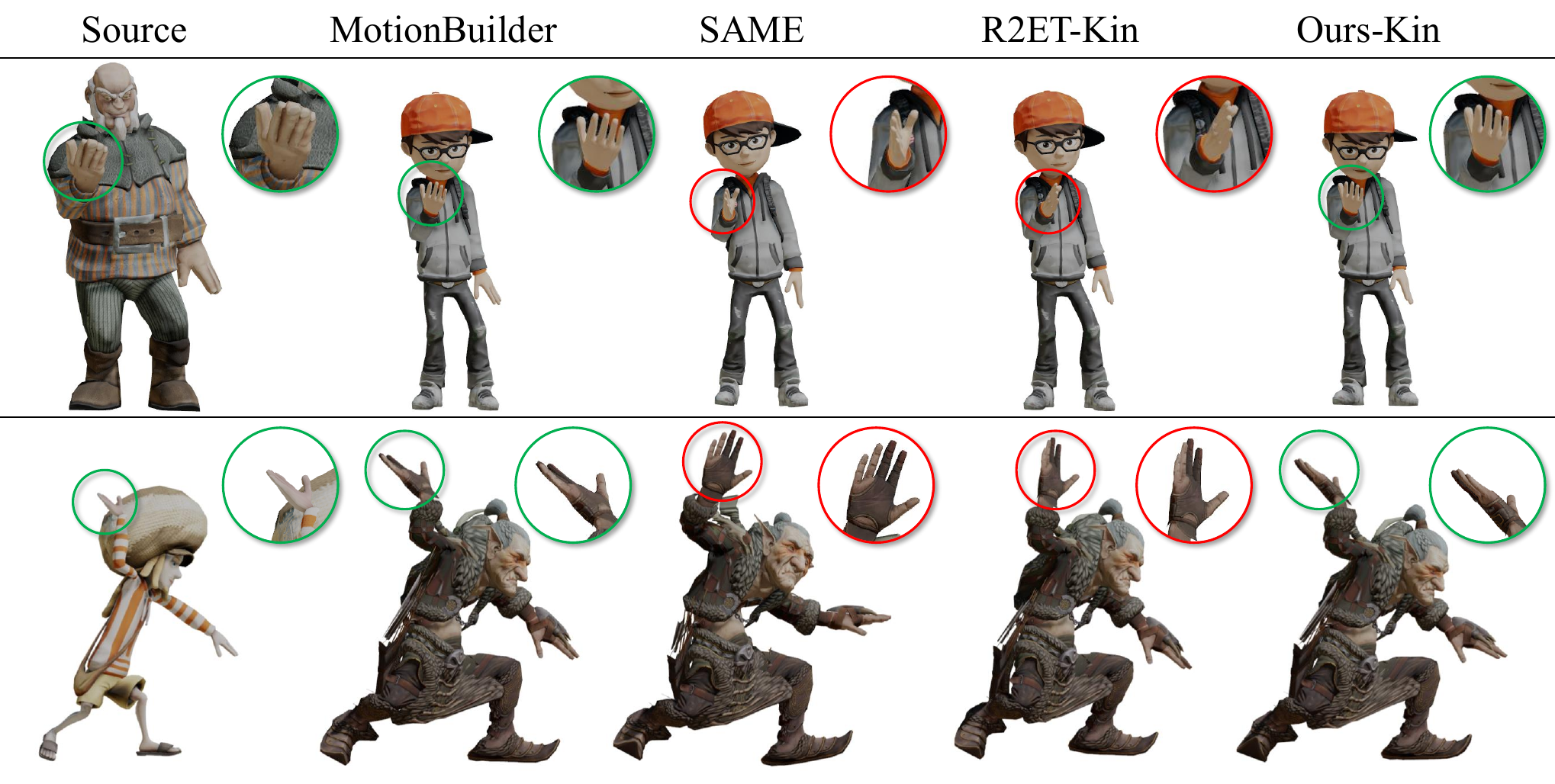}
    \vspace{-1em}
    \caption{Qualitative results on kinematic motion retargeting.}
   \label{fig:kin_qual}
    \vspace{-0.5em}
\end{figure*}

%% file: fig/geo_qual.tex
\begin{figure*}[t]
    \centering
    \includegraphics[width=\linewidth]{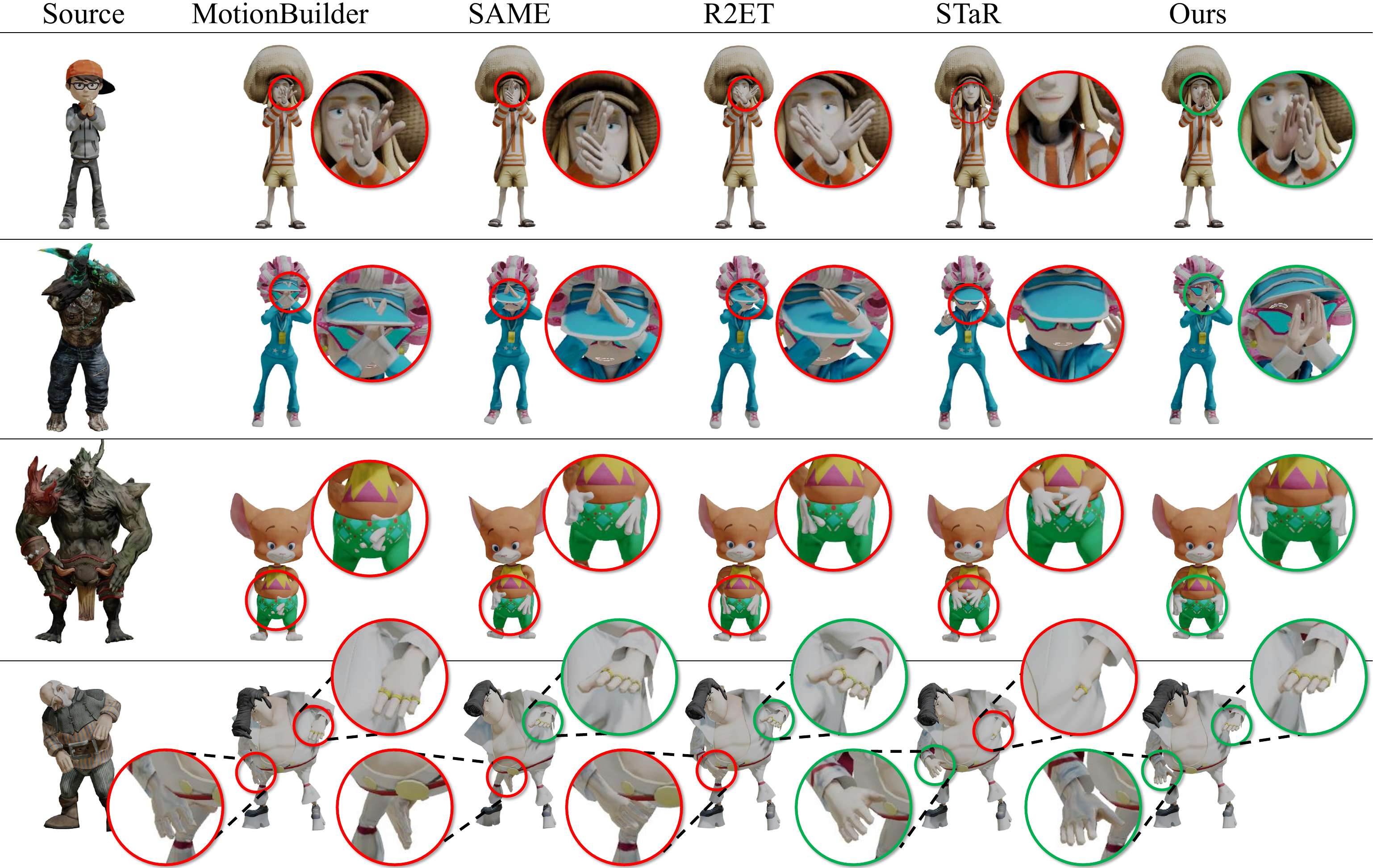}
    \vspace{-2em}
    \caption{Qualitative results on geometry-aware motion retargeting.}
   \label{fig:geo_qual}
\end{figure*}

%% file: tab/ablation_geometry.tex
\begin{table*}[t]
\centering
\caption{Ablation study on geometry-aware refinement in the fixed structure setting. \textbf{Bold} indicates the best result, while SC and UC represent seen character and unseen character, respectively.}
\vspace{-1em}
\resizebox{\linewidth}{!}
{
\begin{tabular}{l|cccccc|cccccc}
    \toprule
    \multirow{2}{*}{Methods}
    & \multicolumn{6}{c|}{Fixed-SC}
    & \multicolumn{6}{c}{Fixed-UC} \\
    & JR~$\downarrow$ & RT~$\downarrow$ & JP~$\downarrow$ & FS~$\downarrow$ & PR~$\downarrow$ & PD~$\downarrow$ 
    & JR~$\downarrow$ & RT~$\downarrow$ & JP~$\downarrow$ & FS~$\downarrow$ & PR~$\downarrow$ & PD~$\downarrow$ \\
    
    \midrule
    (a) Static
    & 0.1651 & 5.5897 & 8.4156 & 0.0342 & 0.7934 & 1.4944
    & 0.1635 & 8.1692 & 11.8455 & 0.0309 & 0.2258 & 1.1307 \\
    
    (b) Jacobian
    & 0.1677 & 5.1947 & 8.0185 & \textbf{0.0331} & 0.9262 & 1.4983
    & \textbf{0.1507} & \textbf{5.5421} & \textbf{8.0256} & \textbf{0.0257} & 0.2328 & 1.1410 \\
    
    (c) Static+Jacobian
    & 0.1649 & 5.1511 & 8.1620 & 0.0365 & 0.6624 & 1.4504
    & 0.1648 & 8.8790 & 12.7437 & 0.0296 & 0.1897 & 1.1028 \\
    
    (d) Static+Skinning
    & \textbf{0.1628} & 5.4081 & 8.1972 & 0.0344 & 0.6767 & 1.4460
    & 0.1535 & 6.7474 & 9.4648 & 0.0312 & 0.1956 & 1.0431 \\
    
    \rowcolor{black!16}
    (e) Static+Skinning+Jacobian~(Ours)
    & 0.1633 & \textbf{4.8692} & \textbf{7.6974} & 0.0365 & \textbf{0.6434} & \textbf{1.4311}
    & 0.1556 & 7.4929 & 10.4406 & 0.0393 & \textbf{0.1674} & \textbf{1.0191} \\
    \bottomrule
\end{tabular}
}
\label{tab:ablation_geometry}
\end{table*}

%% file: fig/ablation_geo.tex
\begin{figure*}[ht]
    \centering
    \includegraphics[width=\linewidth]{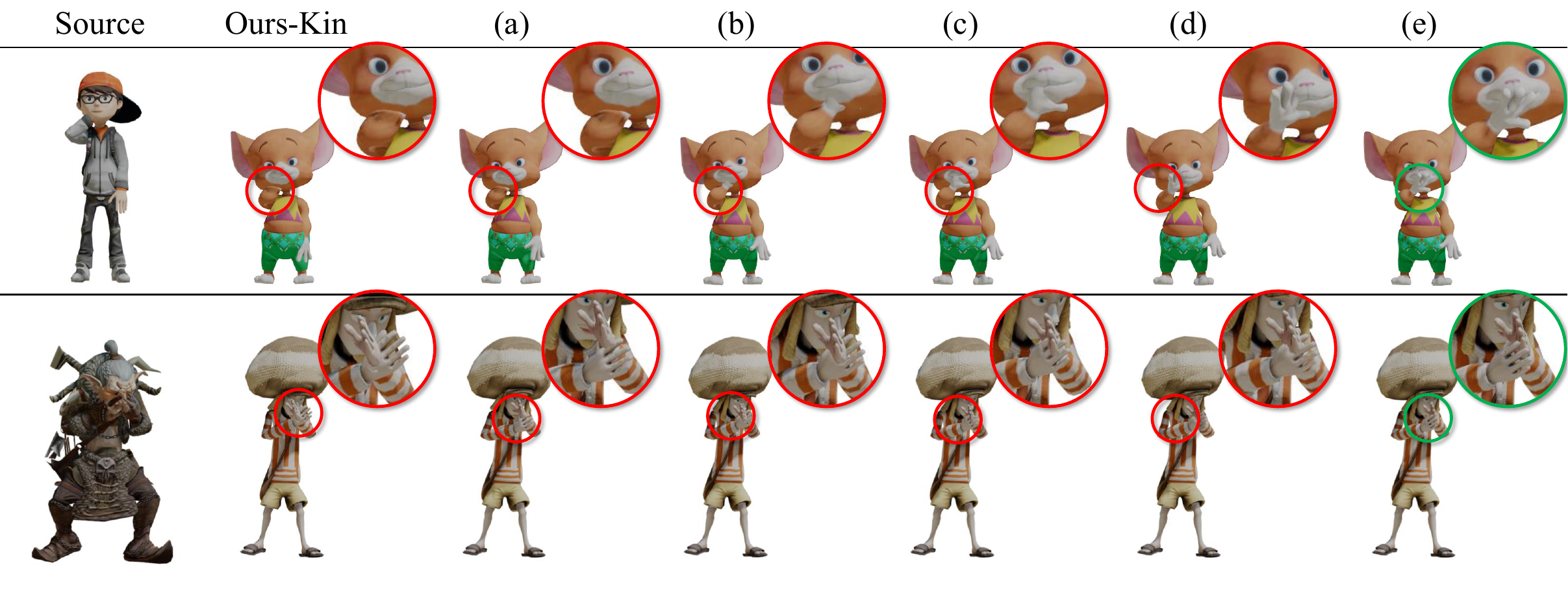}
    \vspace{-2em}
    \caption{Ablation study results on geometry-aware motion retargeting.}
   \label{fig:ablation_geo}
\end{figure*}

%% file: tab/ablation_kin_embedding.tex
\begin{table*}[t]
\centering
\caption
{
    Ablation study on the learnable kinematic motion embedding. \textbf{Bold} indicates the best result, while SC and UC represent seen character and unseen character, respectively.
}
\vspace{-1em}
\resizebox{\linewidth}{!}
{
    \begin{tabular}{c | cccc | cccc | cccc | cccc}
        \toprule
        \multirow{2}{*}{Methods}
        & \multicolumn{4}{c|}{Fixed-SC}
        & \multicolumn{4}{c|}{Fixed-UC}
        & \multicolumn{4}{c|}{Arbitrary-SC}
        & \multicolumn{4}{c}{Arbitrary-UC} \\
        & JR~$\downarrow$ & RT~$\downarrow$ & JP~$\downarrow$ & FS~$\downarrow$ 
        & JR~$\downarrow$ & RT~$\downarrow$ & JP~$\downarrow$ & FS~$\downarrow$ 
        & JR~$\downarrow$ & RT~$\downarrow$ & JP~$\downarrow$ & FS~$\downarrow$ 
        & JR~$\downarrow$ & RT~$\downarrow$ & JP~$\downarrow$ & FS~$\downarrow$  \\

        \midrule
        
        Pooling
        & \textbf{0.1537} & 3.7027 & 5.6877 & \textbf{0.0037}
        & \textbf{0.1549} & 3.9076 & 5.8520 & 0.0040
        & \textbf{0.1536} & 3.4938 & 5.1919 & 0.0047
        & \textbf{0.1537} & 3.0402 & 4.8084 & \textbf{0.0076} \\
        
        \rowcolor{black!16}
        Learnable~(Ours)
        & 0.1561 & \textbf{2.9568} & \textbf{5.2375} & 0.0063
        & 0.1565 & \textbf{2.9869} & \textbf{5.0958} & \textbf{0.0026}
        & 0.1570 & \textbf{2.9294} & \textbf{4.8443} & \textbf{0.0036}
        & 0.1579 & \textbf{2.5759} & \textbf{4.5406} & 0.0089 \\
        
        \bottomrule
    \end{tabular}
}
\label{tab:ablation_kin_embedding}
\end{table*}

%% file: tab/ablation_contrastive.tex
    
    

\begin{table*}[t]
\centering
\caption{Ablation study on the embedding objective. \textbf{Bold} indicates the best result, while SC and UC represent seen character and unseen character, respectively.}
\vspace{-1em}
\resizebox{\linewidth}{!}
{
\begin{tabular}{l|cccc|cccc|cccc|cccc}
    \toprule
    \multirow{2}{*}{Methods}
    & \multicolumn{4}{c|}{Fixed-SC}
    & \multicolumn{4}{c|}{Fixed-UC}
    & \multicolumn{4}{c|}{Arbitrary-SC}
    & \multicolumn{4}{c}{Arbitrary-UC} \\
    & JR~$\downarrow$ & RT~$\downarrow$ & JP~$\downarrow$ & FS~$\downarrow$
    & JR~$\downarrow$ & RT~$\downarrow$ & JP~$\downarrow$ & FS~$\downarrow$
    & JR~$\downarrow$ & RT~$\downarrow$ & JP~$\downarrow$ & FS~$\downarrow$
    & JR~$\downarrow$ & RT~$\downarrow$ & JP~$\downarrow$ & FS~$\downarrow$ \\
    \midrule
    (a) No embedding loss
    & 0.1615 & 3.0295 & 5.6182 & \textbf{0.0055} & 0.1619 & 2.9659 & 5.6135 & \textbf{0.0025} & 0.1608 & 3.0614 & 5.2153 & \textbf{0.0029} & 0.1623 & 2.8180 & 5.0315 & \textbf{0.0054} \\
    
    (b) Pull-only
    & 0.1685 & \textbf{2.9380} & 5.5053 & 0.0071 & 0.1725 & \textbf{2.6651} & \textbf{5.0397} & \textbf{0.0025} & 0.1706 & \textbf{2.6767} & 4.8767 & 0.0038 & 0.1680 & \textbf{2.5013} & 4.6826 & 0.0094 \\
    
    \rowcolor{black!16}
    (c) Contrastive~(Ours)
    & \textbf{0.1561} & 2.9568 & \textbf{5.2375} & 0.0063 & \textbf{0.1565} & 2.9869 & 5.0958 & 0.0026 & \textbf{0.1570} & 2.9294 & \textbf{4.8443} & 0.0036 & \textbf{0.1579} & 2.5759 & \textbf{4.5406} & 0.0089 \\
    \bottomrule
\end{tabular}
}
\label{tab:ablation_contrastive}
\end{table*}

%% file: tab/optim.tex
\begin{table*}[t]
\centering
\caption{Comparison with optimization-based methods in the fixed structure setting. \textbf{Bold} indicates the best result, while SC and UC represent seen character and unseen character, respectively.}
\vspace{-1em}
\resizebox{\linewidth}{!}
{
\begin{tabular}{l|ccccccc|ccccccc}
    \toprule
    \multirow{2}{*}{Methods}
    & \multicolumn{7}{c|}{Fixed-SC}
    & \multicolumn{7}{c}{Fixed-UC} \\
    & JR~$\downarrow$ & RT~$\downarrow$ & JP~$\downarrow$ & FS~$\downarrow$ & PR~$\downarrow$ & PD~$\downarrow$ & IT~$\downarrow$ 
    & JR~$\downarrow$ & RT~$\downarrow$ & JP~$\downarrow$ & FS~$\downarrow$ & PR~$\downarrow$ & PD~$\downarrow$ & IT~$\downarrow$ \\
    
    \midrule
    Optimization
    & \textbf{0.1251} & \textbf{2.4697} & \textbf{4.6811} & \textbf{0.0123} & \textbf{0.6031} & \textbf{1.1791} & 15.6788
    & \textbf{0.1160} & \textbf{6.3089} & \textbf{8.2005} & \textbf{0.0227} & \textbf{0.1265} & \textbf{0.7556} & 98.7563 \\
    
    
    \rowcolor{black!16}
    Learning~(Ours)
    & 0.1633 & 4.8692 & 7.6974 & 0.0365 & 0.6434 & 1.4311 & \textbf{0.1646}
    & 0.1556 & 7.4929 & 10.4406 & 0.0393 & 0.1674 & 1.0191 & \textbf{0.9157} \\
    \bottomrule
\end{tabular}
}
\label{tab:optim}
\end{table*}

%% file: fig/optim.tex
\begin{figure}[t]
    \centering
    \includegraphics[width=\linewidth]{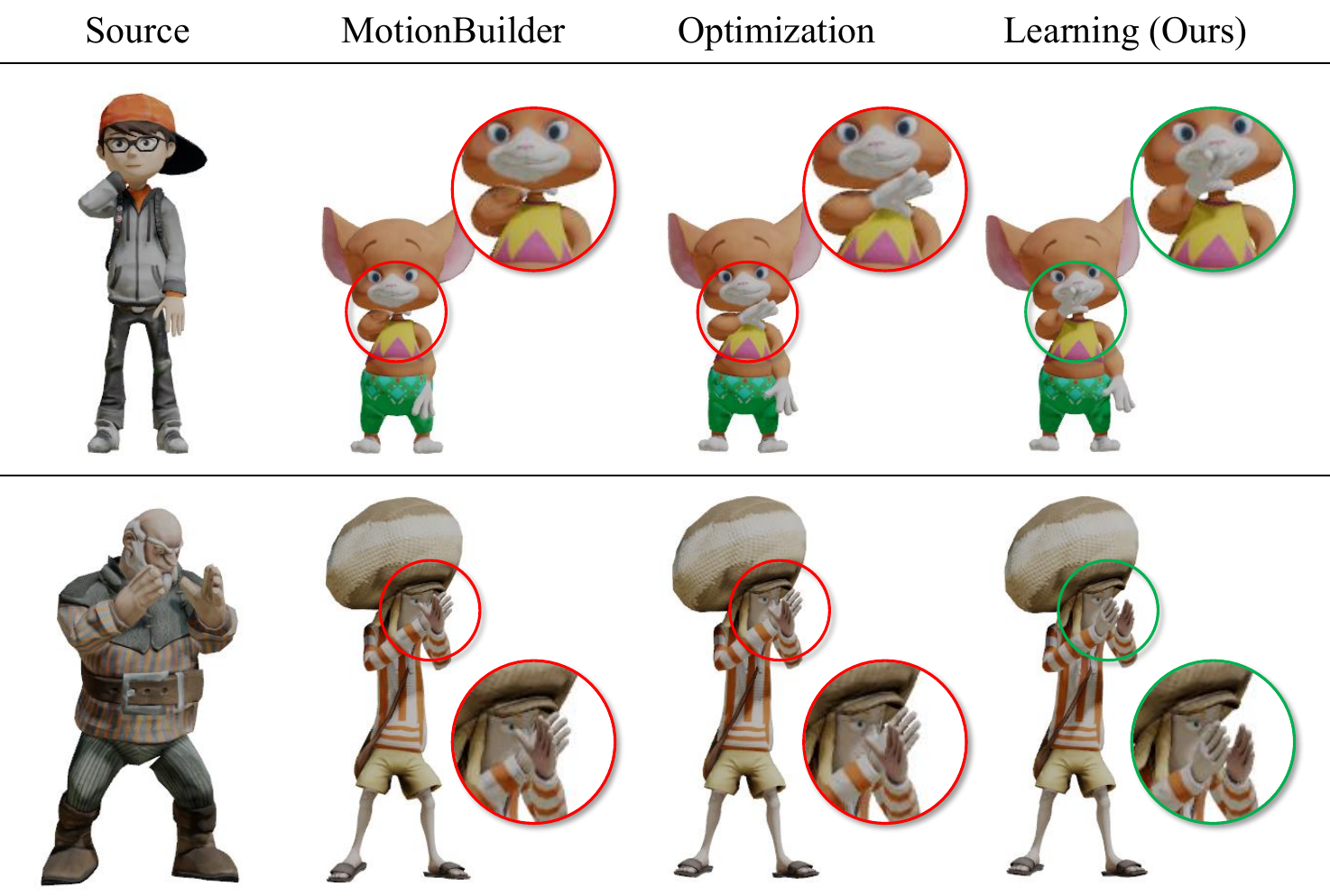}
    \caption{Qualitative comparison with post-hoc optimization.}
   \label{fig:optim}
\end{figure}

%% file: tab/ablation_decim.tex
\begin{table}[t]
\centering
\caption{Ablation study on mesh resolution sensitivity. SC and UC represent seen character and unseen character, respectively.}
\vspace{-0.5em}
\resizebox{\columnwidth}{!}{
\begin{tabular}{lcccccc}
    \toprule
    \multicolumn{7}{c}{Fixed-SC} \\
    \midrule
    Methods
    & JR~$\downarrow$ & RT~$\downarrow$ & JP~$\downarrow$
    & FS~$\downarrow$ & PR~$\downarrow$ & PD~$\downarrow$ \\
    \midrule
    Raw
    & 0.1633 & 4.8692 & 7.6974 & 0.0365 & 0.6434 & 1.4311 \\

    Decim (5000)
    & 0.1633 & 5.0052 & 7.7987 & 0.0369 & 0.7009 & 1.4439 \\
    \midrule
    \multicolumn{7}{c}{Fixed-UC} \\
    \midrule
    Methods
    & JR~$\downarrow$ & RT~$\downarrow$ & JP~$\downarrow$
    & FS~$\downarrow$ & PR~$\downarrow$ & PD~$\downarrow$ \\
    \midrule
    Raw
    & 0.1556 & 7.4929 & 10.4406 & 0.0393 & 0.1674 & 1.0191 \\

    Decim (5000)
    & 0.1545 & 5.9629 & 8.8940 & 0.0379 & 0.1858 & 1.0493 \\
    \bottomrule
\end{tabular}
}
\vspace{-1em}
\label{tab:ablation_decim}
\end{table}

%% file: fig/ood_mixamo.tex
\begin{figure*}[t]
    \centering
    \includegraphics[width=\linewidth]{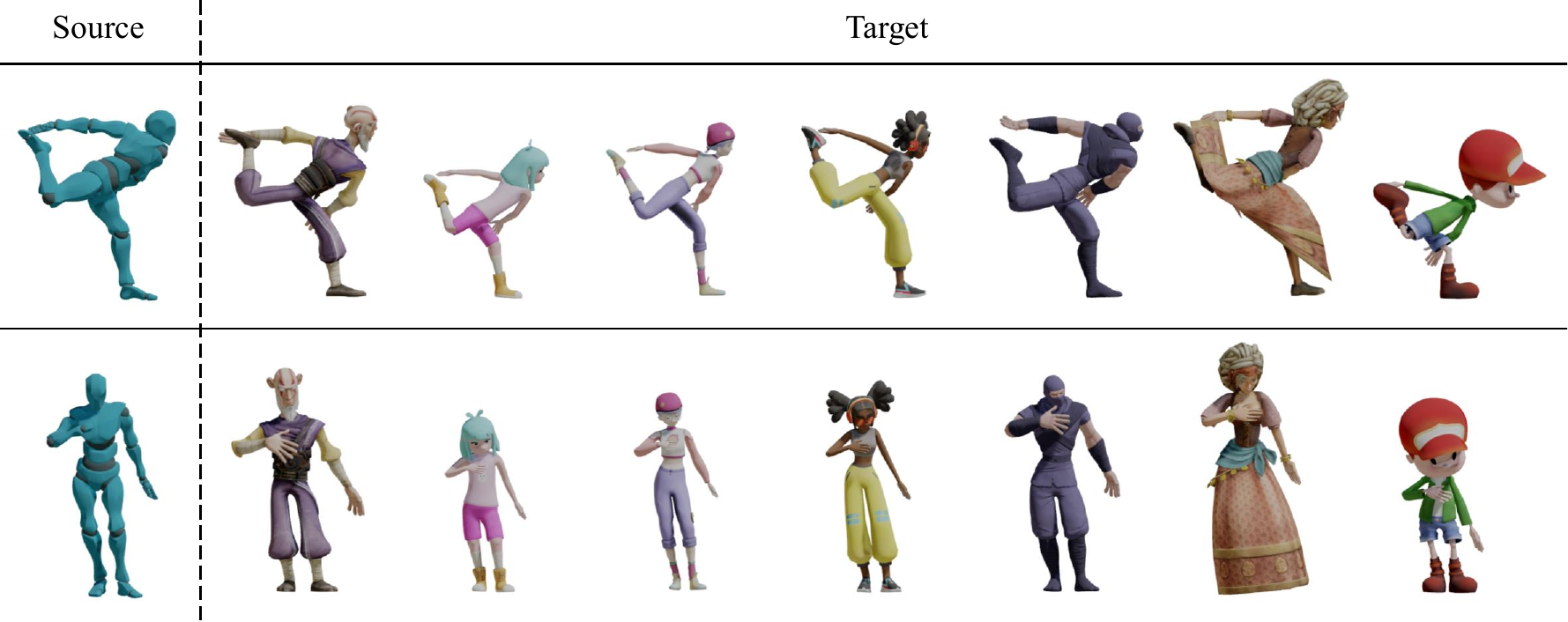}
    \vspace{-2em}
    \caption{Additional qualitative results on several Mixamo characters.}
   \label{fig:ood_mixamo}
   \vspace{-1em}
\end{figure*}

%% file: fig/ood.tex
\begin{figure}[t]
    \centering
    \includegraphics[width=\linewidth]{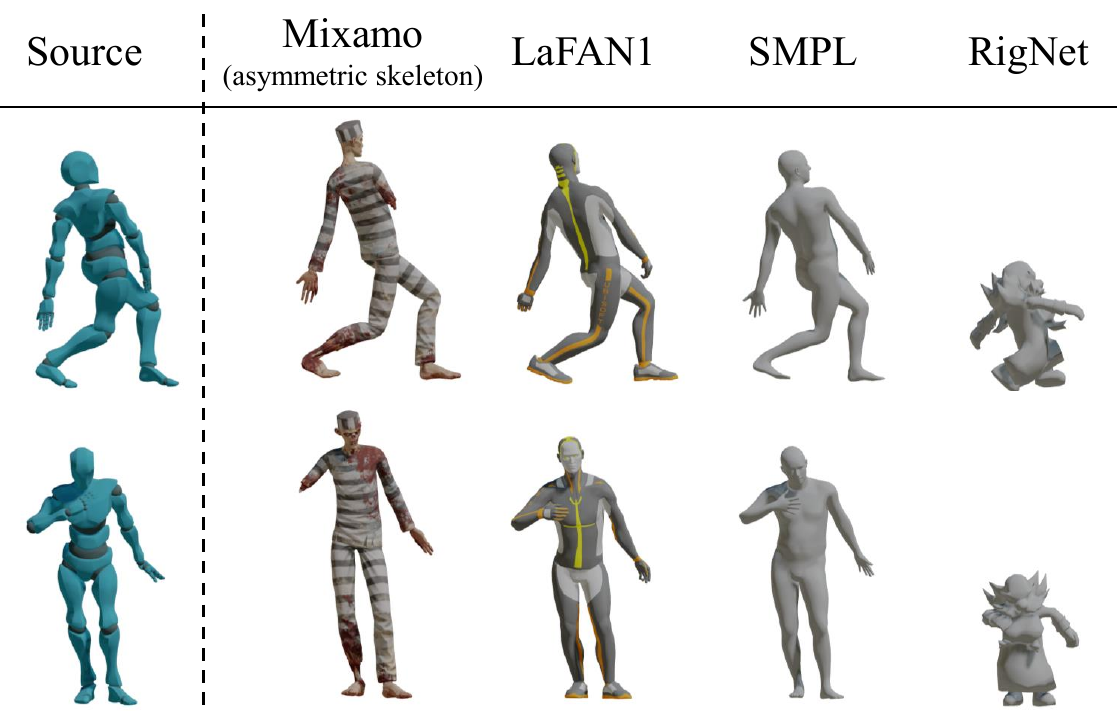}
    \vspace{-2em}
    \caption{Qualitative results on out-of-distribution characters.}
   \label{fig:ood}
\end{figure}

\begin{figure}[t]
    \centering
    \includegraphics[width=0.7\linewidth]{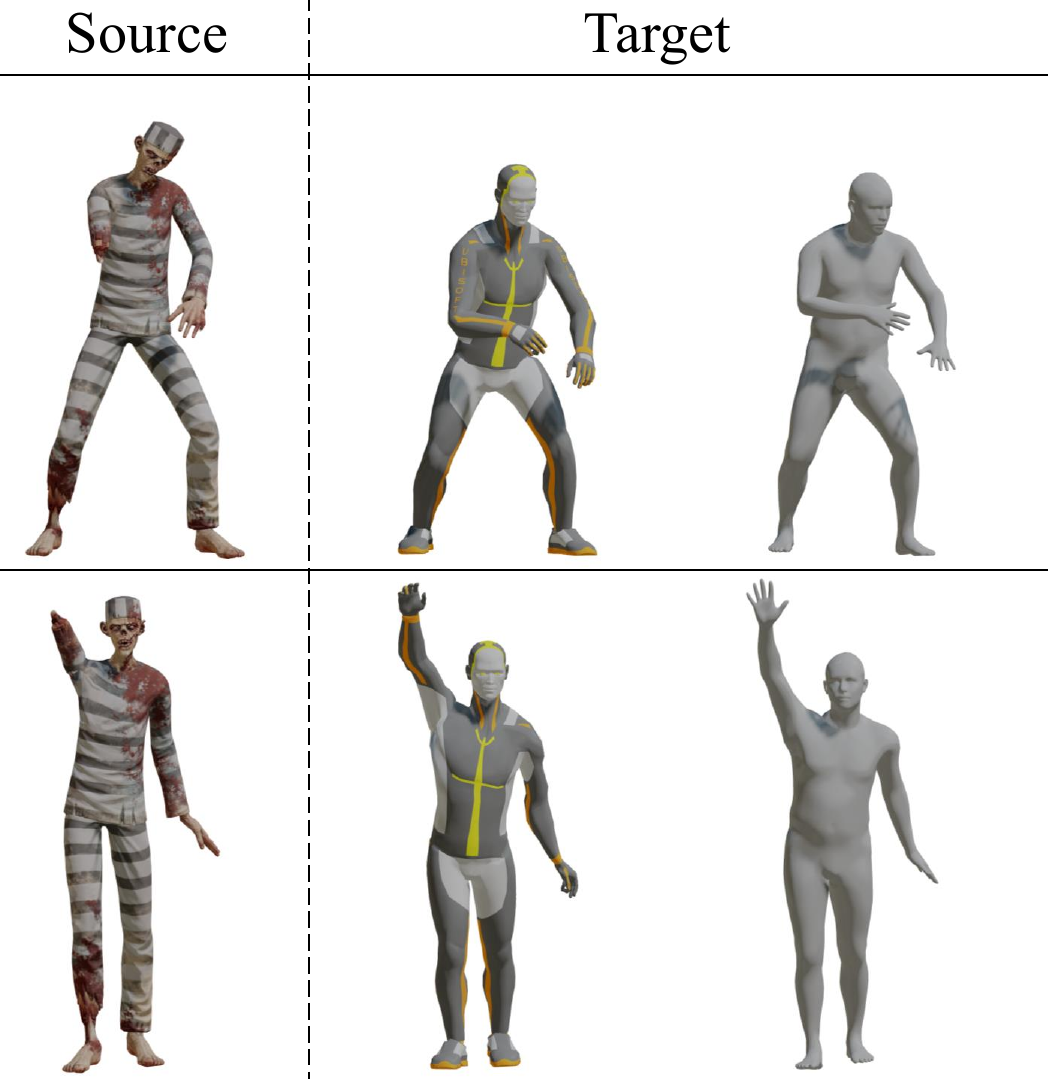}
    \vspace{-1em}
    \caption{Qualitative results of motion retargeting from an asymmetric source skeleton to out-of-distribution target skeletons.}
   \label{fig:ood2}
\end{figure}

%% file: fig/self_pen_recovery.tex
\begin{figure*}[t]
    \centering
    \includegraphics[width=\linewidth]{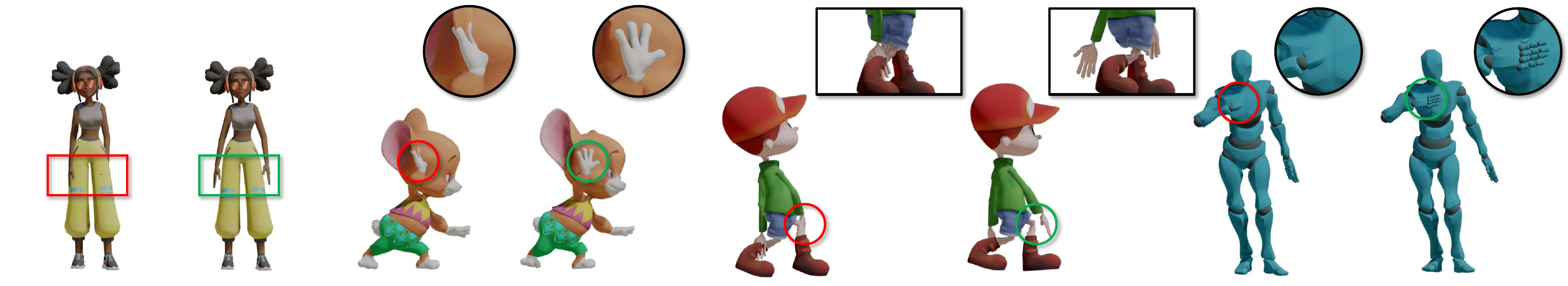}
    \vspace{-2em}
    \caption{Penetration recovery through self-reconstruction.}
   \label{fig:self_pen_recovery}
\end{figure*}

%% file: tex/5_discussion.tex
\section{Discussion}
\label{sec:discussion}
Although our method explicitly observes artifacts from posed character geometry and uses them to improve geometric generalization, it still has several limitations.
First, artifact observation and correction introduce additional computation beyond a standard feed-forward network pass.
In particular, the full pipeline can be executed only after the target-side motion realization, because forward kinematics, skinning, penetration detection, Jacobian computation, and corrective decoding all depend on the decoded target motion and posed geometry.
To make this overhead more transparent, we measured the runtime of each component separately with a 64-frame motion sequence, including target-side kinematic decoding, forward kinematics, skinning-based deformation using the LBS, penetration detection, and Jacobian computation, as summarized in~\cref{tab:limitation_computation}.
Notably, penetration detection was the dominant computational bottleneck, and its cost increased substantially with the number of vertices.
In contrast, the refinement components remained relatively lightweight, indicating that the main overhead comes from explicit mesh-level artifact analysis rather than the neural components.
A promising direction for future work is to reduce the cost of artifact observation, for example by learning surrogate models for penetration analysis and motion-to-vertex sensitivity.
Such surrogates could reduce post-decoding overhead while retaining the explicit corrective structure.

\input{tab/limitation_computation}
\input{fig/limitation}

Our method may still exhibit residual artifacts, including unresolved self-penetration, inaccurate poses, and foot sliding.
In particular, it may fail for highly exaggerated skeletal morphologies or strongly out-of-distribution poses that differ substantially from those observed during training, as shown in~\cref{fig:limitation}.
This limitation becomes more pronounced in complex multi-contact scenarios, where the penetration avoidance adversely affects contact preservation.
Moreover, geometry-aware refinement can increase foot sliding as reflected in~\cref{tab:geometry_template_retargeting}, especially for unseen characters, because reducing geometric artifacts may require pose adjustments that conflict with preserving the provisional kinematic motion.
Therefore, future work could expand the coverage of the training data to improve generalization and incorporate contact-aware inverse kinematics~(IK) post-processing to mitigate foot sliding.

Finally, our corrective signal is derived from a first-order local approximation of geometry artifacts from posed character geometry.
Although iterative training allows the network to learn more global and non-linear correction patterns on top of this signal, the initial cue itself remains local and linear, and it does not explicitly model coordinated adjustments across multiple body parts.
Therefore, future work could explore richer corrective formulations beyond first-order sensitivity, so that the model can better capture globally coherent body movements and more complex correction landscapes.

%% file: tab/limitation_computation.tex
\begin{table*}[t]
\centering
\caption{
    Runtime analysis of geometry-aware correction components with different numbers of mesh vertices.
    We report execution time in seconds, with the percentage of total runtime in parentheses.
}
\label{tab:limitation_computation}
        \begin{tabular}{c|cccccc|c}
            \toprule
            $N_V$ & \makecell{Kinematic \\ Decoding} & FK+LBS & \makecell{Geometry \\ Enrichment} & \makecell{Penetration \\ Detection} & \makecell{Jacobian-based \\ Correction} & \makecell{Corrective \\ Decoding} & Total \\
            \midrule
            5750  & 0.0010~(3.80\%) & 0.0057~(21.67\%) & 0.0002~(0.76\%) & 0.0146~(55.51\%) & 0.0032~(12.17\%) & 0.0016~(6.08\%) & 0.0263 \\
            7171  & 0.0014~(2.16\%) & 0.0086~(13.29\%) & 0.0007~(1.08\%) & 0.0474~(73.26\%) & 0.0048~(7.42\%) & 0.0018~(2.78\%) & 0.0647 \\
            24798 & 0.0010~(0.19\%) & 0.0059~(1.11\%) & 0.0008~(0.15\%) & 0.5153~(96.84\%) & 0.0076~(1.43\%) & 0.0015~(0.28\%) & 0.5321 \\
        \bottomrule
        \end{tabular}
\end{table*}

%% file: fig/limitation.tex
\begin{figure}[t]
    \centering
    \includegraphics[width=1.0\linewidth]{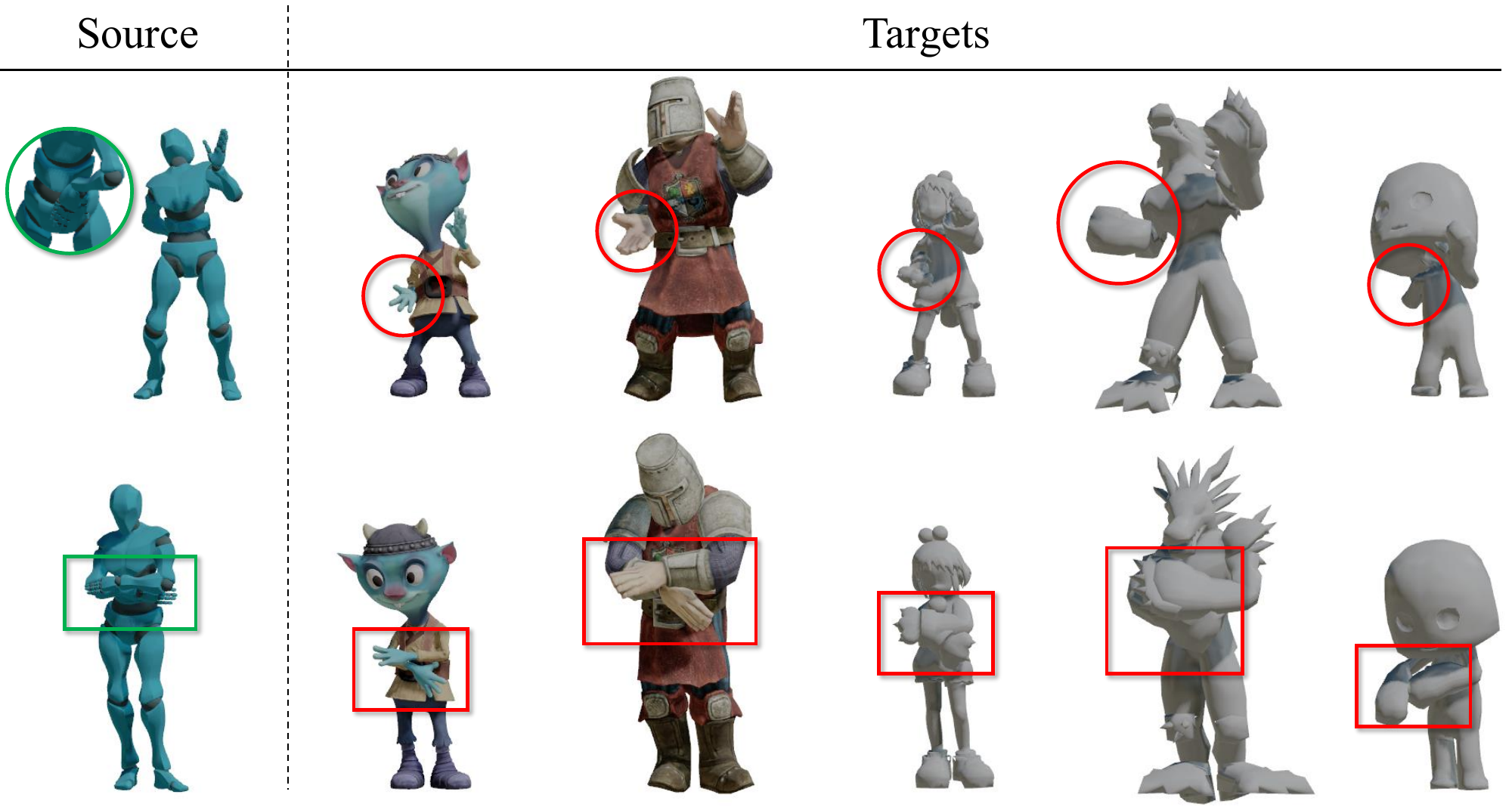}
    \caption{Examples of failure cases. Gray target characters have exaggerated skeletal proportions and mesh geometries.}
   \label{fig:limitation}
\end{figure}

%% file: tex/6_conclusion.tex
\section{Conclusion}
\label{sec:conclusion}
In this work, we presented a learning-based geometry-aware motion retargeting framework that preserves the flexibility of neural retargeting while explicitly addressing target-side geometric artifacts.
Instead of relying on a single network to solve the entire geometry-aware retargeting problem implicitly, we decomposed the process into distinct components with separate responsibilities: kinematic retargeting, target-side deformation through skinning, penetration observation, artifact-driven corrective cue derivation, and geometry-aware decoding.
We first trained a transformer-based kinematic autoencoder to learn a skeleton-agnostic motion embedding that serves as a transferable motion prior across diverse characters.
Building on this prior, we introduced an artifact-driven geometry-aware motion refinement framework that converts posed target-side artifacts into explicit refinement cues using a motion-to-vertex Jacobian.
We further used skinning weights as a mediator between geometry and skeletons, enabling character-specific geometric information to be incorporated in a joint-aligned manner while preserving dimensional flexibility.
Experiments demonstrated that our method improves both kinematic retargeting accuracy and geometric plausibility across diverse skeletons, body shapes, and diverse target characters with different skeleton and mesh structures.

%% file: tex/9_supp.tex
\section{Implementation Details}
\subsection{Network Architecture}
The encoder and decoder of the transformer-based kinematic autoencoder share the same architecture consisting of 3 transformer encoder layers.
Following the pre-layer normalization transformer design~\cite{xiong2020layer}, each layer is composed of layer normalization, multi-head self-attention, and a feed-forward network with Gaussian error linear unit~(GELU) activation~\cite{hendrycks2016gaussian}, where residual connections are applied after the self-attention and feed-forward blocks.
We set the number of attention heads, hidden dimension, feed-forward dimension, motion embedding dimension to 8, 256, 1024, and 32, respectively.
The geometry-aware refinement components, including $\phi_\mathrm{corr}(\cdot)$, $\phi_\mathrm{geo}(\cdot)$, and $\phi_\mathrm{enr}(\cdot)$ are implemented as 2-layer multi-layer perceptrons~(MLPs) with GELU activation bewteen the two linear layers.

\subsection{Penetration Computation}
Our penetration detection is performed in a part-wise nearest-neighbor search manner.
For each sample in a mini-batch, we define query vertices from limb regions and reference vertices from the remaining body parts, and evaluate penetration only between these query--reference pairs.
To preserve parallelism even when each character contains a different number of valid vertices, we first mask invalid vertices and then pack the selected query and reference vertices into contiguous tensors with batch-wise padding to the maximum valid count in the batch.
This allows all pair-wise nearest-neighbor queries to be computed in parallel using batched tensor operations, while still supporting characters with different mesh resolutions and different numbers of valid vertices.
To further reduce the search space, we prune reference vertices by an expanded query-side bounding box before computing nearest neighbors.

Given the nearest reference vertex for each query vertex, we define penetration using the displacement from the query to the reference together with the reference normal.
A query vertex is considered penetrating only when the displacement has a positive projection onto the reference normal, the nearest-neighbor distance is smaller than a threshold, and the normal similarity is below a threshold.
In practice, the distance filtering introduces a trade-off: it may increase false negatives by missing some long-range penetrations, but it effectively suppresses false positives caused by unrelated distant correspondences, which is often observed in convex hull-based voxelized signed distance field-based penetration detection~\cite{zhang2023r2et}.
We found this trade-off beneficial for neural network training, because false positives introduce supervision that is inconsistent with the actual objective of penetration correction.
In particular, they force the model to explain and correct penetrations that do not truly exist, which makes the corrective target noisier and destabilizes learning.
In contrast, false negatives only omit some valid correction signals, and are therefore less disruptive in our setting.
Because the detector is applied repeatedly over the course of training, the model can progressively suppress stronger and more reliable penetrations first, after which weaker or previously missed cases can become exposed and corrected as well.
As a result, the detector provides a cleaner and more stable artifact signal for training the refinement module.

\subsection{Training and Evaluation}
\paragraph{Loss Weights}
We set $\lambda_q$, $\lambda_p$, $\lambda_r$, $\lambda_\mathrm{vel}$, $\lambda_\mathrm{jerk}$, $\lambda_c$, $\lambda_{cv}$, $\lambda_\mathrm{slide}$, $\lambda_\mathrm{gp}$, $\lambda_z$, $\lambda_\mathrm{pen}$, and $\lambda_\mathrm{sparse}$ to 5.0, 0.01, 10.0, 1.0, 0.2, 1.0, 6.0, 6.0, 0.1, 1.0, 0.1, and 0.1, respectively.

\paragraph{Dataset}
To train the geometry-aware refinement module, we used 7 skinned characters from Mixamo: \textit{Aj}, \textit{BigVegas}, \textit{Goblin}, \textit{Kaya}, \textit{Mousey}, \textit{PeasantMan}, and \textit{Warrok}.
For quantitative evaluation, we used 4 skinned characters from Mixamo: \textit{Claire}, \textit{Mutant}, \textit{Ortiz}, and \textit{SportyGranny}.
For additional qualitative results, we used 8 more Mixamo characters: \textit{Abe}, \textit{Amy}, \textit{CastleGuard-01}, \textit{Jackie}, \textit{Michelle}, \textit{Ninja}, \textit{PeasantGirl}, and \textit{YBot}.
We additionally included out-of-distribution characters from LaFAN1, SMPL, and RigNet-v1~(sample 2977)~\cite{xu2020rignet}.


\section{Applications}
Structured motion embeddings have been shown to support diverse downstream tasks beyond retargeting~\cite{lee2023same}.
Similarly, our kinematic transformer can also be applied to several applications, because it is trained with a contrastive embedding loss that encourages the latent space to preserve motion semantics while remaining discriminative across different motions.
In this section, we show that our method preserves this useful property across diverse downstream tasks, including classification, arithmetic operations, and motion control.

\subsection{Classification}
\input{tab/classification}
Using the motion embedding extracted from the encoder, we train a motion classifier that maps an embedding to an output label.
Specifically, the classifier is implemented as a single-layer encoder-decoder block transformer~\cite{vaswani2017attention}.
For training and evaluation, we used the CMU motion database~\cite{cmu2006}, which contains 276 motion clips from 40 subjects spanning 30 categories.
Following the same protocol as SAME, we construct a validation set of 60 clips and use the remaining 216 clips for training.

As shown in~\cref{tab:classification}, the classifier trained on our embedding achieved higher accuracy than the one trained on SAME.
In particular, SAME reported an accuracy of 0.950~(i.e. 57 out of 60), whereas our embedding achieved 0.983~(i.e. 59 out of 60) on the same validation set.
We attribute this improvement to the more discriminative structure of our embedding space, which is encouraged by the contrastive objective and the transformer-based global aggregation.
These results show that the proposed embedding is not only effective for retargeting, but also serves as a strong representation for downstream motion understanding tasks.

\subsection{Arithmetic Operations}
\input{fig/arith}
Because the motion embedding space is structured such that semantically similar motions are placed nearby, our kinematic model supports arithmetic operations in the latent space.
Given three motions $M^{1:T}_a$, $M^{1:T}_b$, and $M^{1:T}_c$, we extract their corresponding embeddings $z^{1:T}_a$, $z^{1:T}_b$, and $z^{1:T}_c$.
Subsequently, we decode a linear combination of these embeddings to obtain a semantically modified motion.
For example, as shown in~\cref{fig:arith}, decoding $w\cdot(z^{1:T}_a-z^{1:T}_b)+z^{1:T}_c$ yields a semantically edited motion that removes the semantic component of $M^{1:T}_b$ from $M^{1:T}_a$ while adding semantics of $M^{1:T}_c$, while the scalar $w$ controls the strength of the modification.
These results demonstrate that the learned embedding space is not only useful for retargeting, but also supports intuitive and human-interpretable motion editing.

\subsection{Motion Control}
\input{fig/control}
As shown in~\cref{fig:control}, our method also supports interactive character control through motion matching~\cite{clavet2016motion} within the shared motion embedding.
Using a database that combines heterogeneous motion categories, including locomotion, crouching, and kicking, the controller enables smooth transitions across these distinct motion types while remaining in a unified motion representation.
It further supports on-the-fly character switching, showing that the learned embedding is useful not only for retargeting but also for interactive motion control across diverse characters and downstream tasks.

%% file: tab/classification.tex
\begin{table}[t]
\centering
\caption{Motion classification accuracy on the CMU motion database.}
\label{tab:classification}
\begin{tabular}{lc}
\toprule
Method & Accuracy \\
\midrule
SAME~\cite{lee2023same} & 0.950 \\
Ours & \textbf{0.983} \\
\bottomrule
\end{tabular}
\end{table}

%% file: fig/arith.tex
\begin{figure}[t]
    \centering
    \includegraphics[width=\linewidth]{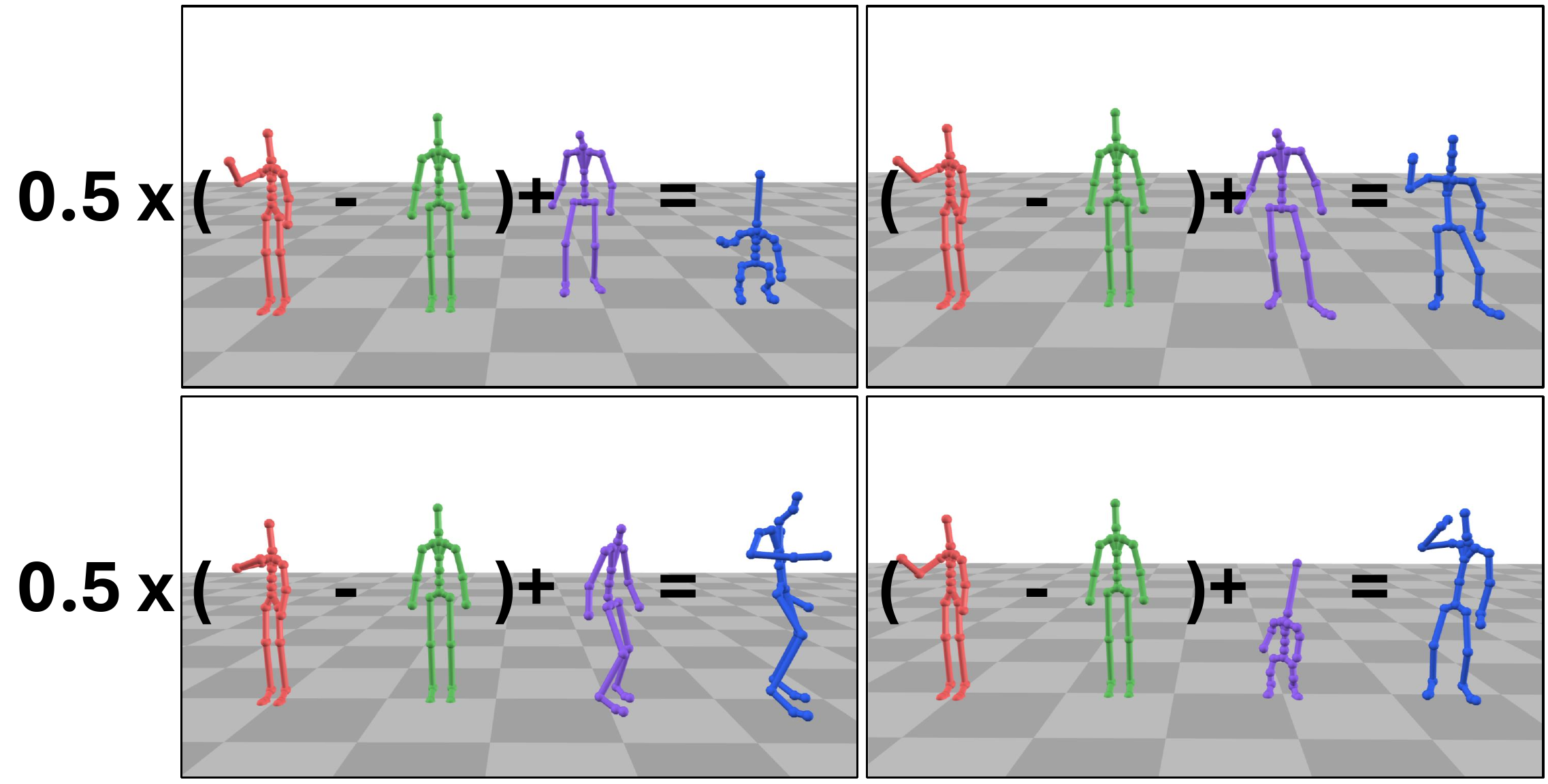}
    \vspace{-2em}
    \caption{Arithmetic operations in the motion embedding space.}
   \label{fig:arith}
\end{figure}

%% file: fig/control.tex
\begin{figure}[t]
    \centering
    \includegraphics[width=\linewidth]{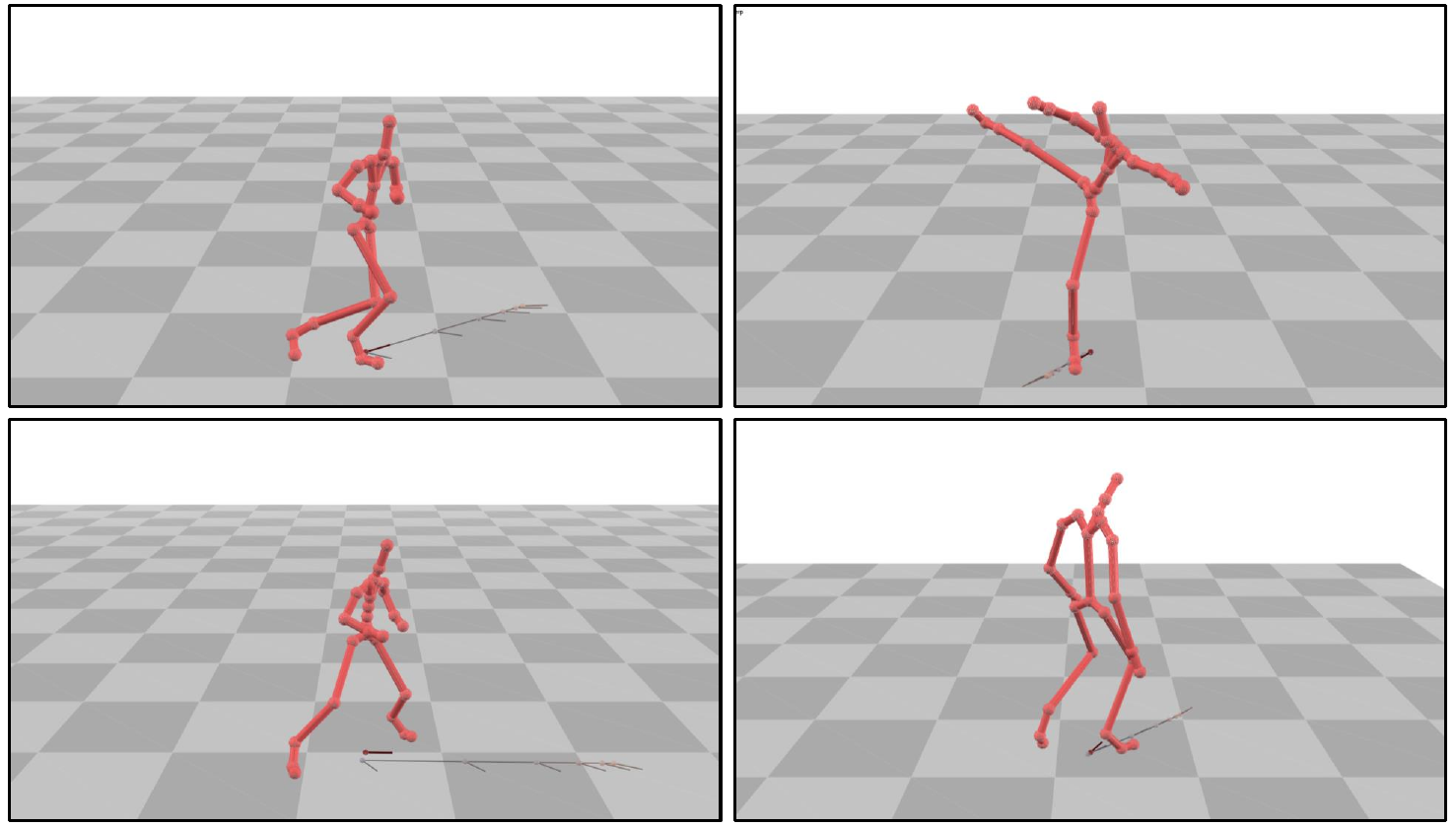}
    \vspace{-2em}
    \caption{Motion control using motion matching within the shared motion embedding space.}
   \label{fig:control}
\end{figure}